\documentstyle[eqsecnum,axodraw,psfig,aps]{revtex}

\def\gsim{\stackrel{>}{\sim}}
\def\lsim{\stackrel{<}{\sim}}
\def\beq{\begin{equation}}
\def\eeq{\end{equation}}
\def\ba{\begin{array}}
\def\ea{\end{array}}

\def\half{\frac{1}{2}}

\def\egzk{E_{GZK}}

\begin{document}
\input{epsf}
\begin{flushright}
submitted to Astro. Part. Phys.\\
preprint VAND--TH--99--12\\
%

\end{flushright}

\begin{centering}

{\large{\bf Signatures for a Cosmic Flux\\
            of Magnetic Monopoles}}
\vspace{1cm}\\

{\large Stuart D. Wick$^1$, Thomas W. Kephart$^1$, Thomas J. Weiler$^1$\\
and Peter L. Biermann$^2$}\\

\vspace{0.5cm}

{$^1$ \it Department of Physics \& Astronomy, Vanderbilt University,\\
        Nashville, TN 37235}\\

{$^2$ \it Max Planck Institute for Radioastronomy,\\
        D-53010 Bonn, Germany}\\

\end{centering}
\vspace{1cm}
\begin{abstract}

Any early Universe phase transition
occurring after inflation has the potential
to populate the Universe with relic magnetic monopoles.
Observations of galactic magnetic fields,
as well as observations matched with models
for extragalactic magnetic fields,
lead to the conclusion that monopoles of
mass $\lsim 10^{15}$ GeV are accelerated
in these fields to relativistic velocities.
We explore the possible signatures
of a cosmic flux of relativistic monopoles impinging on the earth.
The electromagnetically-induced
signatures of monopoles are reliable.
The hadronically-induced signatures are highly model-dependent.
Among our findings are
(i) the electromagnetic energy losses of monopoles
continuously initiate a protracted shower of small intensity;
(ii) monopoles may traverse the earth's diameter,
making them a probe of the earth's interior structure;
(iii)  in addition to the direct monopole Cherenkov
signal presently employed, a very attractive
search strategy for
monopoles is detection of their coherent radio-Cherenkov signal
produced by the charge-excess in the $e^+$--$e^-$ shower ---
in fact, Cherenkov-detectors
have the potential to discover a monopole flux
(or limit it) several orders of magnitude
below the theoretical Parker limit
of $10^{-15}$/$\rm{cm}^2$/s/sr;
(iv) it is conceivable (but not compelling)
that bound states of colored monopoles
may be the primary particles initiating the air
showers observed above the GZK cutoff.

\end{abstract}

\section{Introduction}

Any breaking of a semisimple gauge symmetry
which occurs after inflation and which
will leave unbroken a $U(1)$ symmetry group may produce an observable abundance
of magnetic monopoles.
The initial monopole number density
can be estimated from the type and temperature of the
symmetry breaking phase transition \cite{Kibble}.
From this, one easily derives the present 
day flux of monopoles \cite{LPi}.
The monopole mass is expected to be
$\sim \alpha^{-1}$
times the scale $\Lambda_{SB}$ of the symmetry breaking.
It is noteworthy that the
inferred strength and coherence size
of existing extragalactic magnetic fields suggest
that any monopole with a mass near or less than $10^{14}$ GeV would
have been accelerated in magnetic fields to relativisitic velocities.
On striking matter, such as the earth's atmosphere, these
relativistic monopoles will generate a particle cascade.
It is the purpose of this paper to calculate the shower signatures
of relativistic magnetic monopoles.
An extensive collection of monopole references covering
theoretical investigations and experimental searches
prior to 1998 can be found in \cite{refs}.

The electromagnetic interaction cross--section for
relativistic monopoles is fairly well understood, but
the hadronic cross--section is not.
We calculate in some detail the reliable signatures resulting
from the electromagnetic monopole--matter interaction.
For the hadronic interactions of the monopole, we investigate
various possibilities, and
we present qualitative arguments for the resulting signatures.

On theoretical grounds, the flux
of monopoles that do not catalyze baryon decay 
is limited only by Parker's upper bound
$F_{\rm{P}}\sim 10^{-15}$/cm$^2$/s/sr \cite{Parker}.
We do not consider monopoles which do catalyze baryon
decay, as the flux limits on these render them
not observable.
The Parker bound results from requiring that monopoles not short--circuit
our Galactic magnetic fields faster than the Galactic dynamo can
regenerate them.  The Parker bound is several orders of magnitude
above the observed highest--energy cosmic ray flux.
Thus, existing cosmic ray detectors can
meaningfully search for a monopole flux, and
proposed vast--area cosmic ray detectors
may improve the search sensitivity by many orders of magnitude.

Because of the small inelasticity of monopole interactions, 
the monopole primary will continuously induce an air--shower.  
This is in contrast to nucleon and photon primaries
which transfer nearly all of their energy in the shower initiation.
Thus we expect the monopole shower to be readily distinguished
from non--monopole initiated showers.
However, we also investigate the possibility that
the hadronic interaction of the monopole is sufficiently strong
to produce air--showers with $dE/dx$ comparable to that from nuclear
primaries, in which case existing data would already imply a meaningful
limit on the monopole flux.  One may even speculate that monopoles with
a large $dE/dx$ have been observed, as the primaries producing the
enigmatic showers above the GZK cutoff at
$\sim 5\times 10^{19}$~eV \cite{PorterGoto,KW96}.

The outline of this paper is as follows:
In section (\ref{sec:2}) we discuss the limits on the monopole
mass and number density.  We discuss the variety
of monopoles resulting from possible symmetry breakings,
and examine their natural kinetic energies resulting from acceleration
in large-scale cosmic magnetic fields. In
section (\ref{sec:monostop}) expressions for the energy loss
of relativistic monopoles are presented.
The electromagnetic processes of
ionization, electron--pair production, bremsstrahlung,
and the photonuclear interaction
provide the most reliable expressions.
The hadronic energy loss is also considered.
In section (\ref{sec:monosigns}) we discuss signatures of
relativistic monopoles
arising from their electromagnetic energy loss processes.
We develop a model for the electromagnetic particle shower
induced by relativistic monopoles.  Cherenkov signatures are
also examined, including the coherent radio-Cherenkov signal
resulting from the shower charge excess.  We mention
the possible detection of the coherent Cherenkov signal
in $\rm{km}^{3}$ ice experiments.
It is shown that for certain masses Earth tomography 
with monopoles is practicable.
The monopole's electromagnetic $dE/dx$ in air appears to be too low
to allow their detection via air fluorescence.
However, we argue that some speculative models may allow 
the geometric size of ``baryonic--monopoles'' to grow as a
result of interactions, with the concomitant increase in
their strong cross-section leading to significant air fluorescence.
This possibility is explored in
section (\ref{sec:Mbaryon}) with a simple model for the
growing strong cross--section.
Finally, in section (\ref{sec:concl}) we give concluding remarks.

\section{Monopoles and Magnetic Fields}
\label{sec:2}

\subsection{Monopole Mass and Number Density}

As first shown by 't Hooft and Polyakov \cite{tHooft},
the dynamical content and stability
of magnetic monopoles in particle physics models is
regulated by the patterns of symmetry breaking in these models. A
classification is given in \cite{HK} based on homotopy theory.
For illustration, we consider the simplest case of a semisimple
gauge group  $G$ (i.e., $G$ has no $U(1)$ factors) that breaks to a
subgroup $G'$ that is nonsemisimple (i.e., has at least one $U(1)$
factor which is part of a linear combination making up $U_{EM}(1)$).
At the symmetry-breaking temperature $T_c \sim \langle\phi\rangle$
at which the order parameter $\langle\phi\rangle$ turns on,
monopoles of mass $M \sim T_c/\alpha$ appear, where
$\alpha$ is the fine structure constant at scale $T_c$.
The order parameter can be the VEV
of some scalar field or some bi-fermionic condensate that breaks the symmetry.
We assume that the value of $\langle\phi\rangle$ is at or above the
electroweak (EW) scale $\sim 250 \;\rm{GeV}$ so as to avoid violations of
Standard Model (SM) physics.
This assumption then bounds the monopole mass for our
consideration to be $\gsim 40 \;\rm{TeV}$.
We note that a similar lower mass bound results from (boldly) treating a
classical monopole as a virtual quantum in radiative corrections
to the SM \cite{DeRujula}.

The number density and therefore the flux of monopoles emerging from
a phase transition are determined by the Kibble mechanism \cite{Kibble}.
At the time of the phase transition, roughly one monopole
or antimonopole is produced per correlated volume.
The resulting monopole number density today is
\beq
n_M \sim 10^{-19}\, (T_c/10^{11}{\rm GeV})^3 (l_H/\xi_c)^3\,{\rm cm}^{-3},
\label{density}
\eeq
where $\xi_c$ is the phase transition correlation length,
bounded from above by the horizon size $l_H$ at the time when the
system relaxes to the true broken--symmetry vacuum.
In a second order or weakly first order phase transition,
the correlation length is comparable to the horizon size.
In a strongly first order transition,
the correlation length is considerably smaller than the horizon size.
In section ({\ref{sec:accelerate}) we will show that free
monopoles with
$M\lsim 10^{15}$~GeV are accelerated to relativistic energies by
the cosmic magnetic fields \cite{foot1}. From Eq.(\ref{density}) then,
the general expression for the relativistic monopole
flux may be written \cite{KW96}
\beq
F_M = c\: n_M/4\pi
 \sim 2\times 10^{-4}\, \left(\frac{M}{{10^{15}{\rm GeV}}}\right)^3
\left(\frac{l_H}{\xi_c}\right)^3\,
{\rm cm}^{-2} \;{\rm sec}^{-1}\;{\rm sr}^{-1}\,.
\label{flux}
\eeq

Phenomenologically, the monopole flux is constrained by cosmology and
by astrophysics.
Cosmology requires that the monopole energy density
$\Omega_M$ not be so large as to add observable curvature to the Universe.
From eq.~(\ref{density}) comes an expression for
the monopole mass density today relative to the closure value
\beq
\Omega_M \sim 0.1\, (M/10^{13} {\rm GeV})^4 (l_H/\xi_c)^3.
\label{omega}
\eeq
Monopoles less massive than
$\sim 10^{13} (\xi_c/l_H)^{3/4}$ GeV do not over-curve the Universe.
Astrophysics requires the monopole flux to respect the Parker bound
such that the magnetic field of our galaxy is sustained.
Requiring that the Kibble flux in eq.~(\ref{flux})
be less than the Parker limit
$F_{P}\equiv 10^{-15}/{\rm cm}^2/{\rm sec}/{\rm sr}$,
one derives a combined mass bound \cite{KW96}
\beq
M\lsim 10^{11} (\xi_c/l_H)\, \rm{GeV}\,.
\label{ParKib}
\eeq
This constraint is stronger than the
curvature constraint by about two orders of magnitude
for phase transitions without excessive latent heat.

The energy--density constraint for relativistic monopoles
is of course stronger than that for non--relativistic monopoles
of the same mass.
From eqs.~(\ref{omega}) and (\ref{flux}) we may write for the
{\it relativistic} monopole closure density \cite{KW96}
\beq
\Omega_{RM} \sim \left(\frac{\langle E_M \rangle}{m_{\rm{Pl}}}\right)
   \left(\frac{F_M}{F_P}\right)\;,
\label{RM}
\eeq
where $m_{\rm{Pl}}=1.2 \times 10^{19}$ GeV is the Plank mass.
This shows that a Kibble monopole flux respecting the Parker limit
cannot over-curve the Universe
regardless of the nature of the monopole--creating
phase transition (parameterized by $\xi_c/l_H$), as long as
$\langle E_M \rangle \lsim m_{\rm{Pl}}$.

Minimal $SU(5)$ breaking
gives monopoles of mass $\sim 10^{17} \;\rm{GeV}$.
However, within field theory there exist many possibilities to produce
monopoles with mass below this scale.
For example, there are other chiral $SU(N)$
\cite{SK}
and chiral $O(N)$ models \cite{Hong,Dutta}
with lighter monopoles.
There are also relatively light
monopoles ($M \sim 10^8$~GeV) in the phenomenologically interesting
$SU(15)$ model \cite{FLFK}.
Recently, a new field theoretic possibility has emerged based on
conformal field theory (CFT).
Although originally conceived in the context of type IIB strings
and M-theory compactifications on
higher--dimensional anti-de Sitter space \cite{Mald},
the rules for constructing the four-dimensional CFTs
can be considered independently of any higher
dimensional origins.
These theories are naturally $N=4$ supersymmetric,
but depending on the compactification, can
have reduced $N=0$ or $1$ supersymmetry
and have sensible phenomenology with
a product of $SU(N_i)$ gauge groups.
Three-family non-supersymmetric models have
been constructed with GUT scale as low as a few TeV \cite{PHF}.
These models have light monopoles $\sim 100$ TeV.
Some of these models unify $U_{EM}(1)$ and $SU_{C}(3)$ \cite{PHF2},
while others do not. Thus, monopoles in these models
may or may not have hadronic interactions.

In yet another alternative,
extra dimensions {\sl a la} Kaluza-Klein \cite{Dienes,Dimop} decrease
$M_{GUT}$ if these extra dimensions are not too tightly compactified.
For example, with the introduction of
two extra dimensions of millimeter size, the GUT scale
is lowered to about $100 \;\rm{TeV}$
due to the dramatic change from logarithmic to power law running in the
RG-improved running coupling constant.
This low--scale unification in turn may lead to magnetic monopoles of
mass $\sim 10^4 \;\rm{TeV}$.
A potential concern for monopoles in higher--dimensional theories is
topological instability,
i.e., monopoles in greater than four dimensions may unwind.
However, monopoles made of SM fields confined to our
three--dimensional brane would remain stable.
In fact, there are even richer possibilities for monopoles
in higher--dimensional theories.
New defect solutions may exist in
the larger dimensions, stabilized by topology.
The intersection of these new solutions on the SM brane
could then be monopoles, stabilized by the usual 3D topological argument.

A relatively low GUT scale seems to be necessary to allow light--mass
monopoles.  A low GUT scale will typically
also enable fast proton decay, in violation of experimental lifetime bounds.
However, exceptions abound.
The $SU(15)$ model has no direct gauge induced proton decay.
Proton decay in the MSSM Pati-Salam Model \cite{SK} is
only induced through the hidden sector.
Some of the 4D models listed above have accidental
symmetries
(continuous or discrete) which stabilize the proton, and the
higher dimensional models have additional stabilizing possibilities.

To summarize this section, there are theoretical possibilities
for producing monopoles with mass $\lsim 10^{15}$~GeV, while
avoiding proton decay.
In addition to their electromagnetic properties,
these monopoles may have a strong interaction cross--section.
In the context of the Kibble mechanism for monopole production,
observational bounds on the Universe's curvature constrain the monopole mass
to less than $10^{13}$~GeV. More constraining is
a comparison of the Kibble flux
to the Parker limit which
constrains the monopole mass to less than $10^{11}$~GeV.

We note that in higher dimensional cosmologies
the Kibble flux given in eq.~(\ref{flux}) may be altered.
If the Kibble flux estimate is changed,
then the straightforward Parker upper limit
$F_M\leq 10^{-15}/{\rm cm}^2/{\rm sec}/{\rm sr}$
becomes the only reliable bound on the monopole flux.
Thus, in the spirit of generality, we will let $M$ be a free parameter
and use the Kibble mechanism as a rough guide to $F_M$.
We will, of course, require that $F_M$ obey the Parker limit.
We also will assume that proton decay is avoided in a way that
does not restrict the parameter $M$.

\subsection{Monopole Structure}

The fact that monopoles are topological defects endows them
with a non-trivial internal structure.
Monopoles are classified \cite{Kibble} by their topological winding
in the group manifold.
This topological classification is coarse for GUT monopoles,
which require further classification according to their charges.
Monopole charges are dual to the SM charges.
For example, in a dual theory magnetic charge
and chromomagnetic (or color--magnetic)
charge replace electric charge and chromoelectric
(or color--electric) charge.

Fundamental monopoles can bind to form composite monopoles.
For example, in an $SU(5)$ GUT the fundamental minimally-charged
monopoles are six-fold degenerate.  For an appropriate Higgs
potential there are four other types of stable bound states
formed from the colored fundamental monopoles \cite{Harvey2}.
Remarkably, the spectrum
of bound--state monopoles corresponds almost exactly to the
particle spectrum of a standard model family, which has led 
to an attempt to construct a ``dual standard model''
out of monopoles \cite{Tanmay2}.

We adopt the
nomenclature ``$q$--monopoles'' for those monopoles with
color--magnetic charge, and ``$l$--monopoles'' for those
with only the ordinary $U_{EM}(1)$ magnetic charge.
The possible confinement of q-monopoles via the formation of
$Z_{3}$ color--magnetic ``strings'' has been considered
recently \cite{Goldhaber}.
If such a confinement mechanism is realized, one expected result
would be the formation of color--singlet ``baryonic--monopoles.''
Thus we are led to consider the phenomenology of
two broad classes of monopoles, the $l$-monopoles with $U(1)$
magnetic charge, and the baryonic-monopoles with hadronic interactions
in addition to electromagnetic interactions.

The baryonic--monopole structure is quite different from that of an
$l$--monopole, and as such we expect it to have
a very different cross--section and cosmic ray shower profile.
The internal dynamics of a baryonic-monopole would
approximate that of an ordinary baryon in the QCD string
model, but with $q$--monopoles replacing quarks at the ends of strings.
In particular, the string tensions of the chromomagnetic and chromoelectric
strings are both of order $\Lambda_{QCD}$.
However, the energetics of string-breaking for the two cases are very different.
The chromoelectric string of standard QCD will stretch until
energetics favors breaking and the production of a quark-antiquark pair.
The chromomagnetic string internal to the baryonic-monopole may stretch until
energetics favors breaking and the production of a monopole-antimonopole pair.
From the large mass of the monopole-antimonopole pair,
we infer that an enormous amount of string-stretching without breaking
is possible for a baryonic-monopole.
In sec.~(\ref{sec:Mbaryon}) we discuss the viability of baryonic--monopoles
as candidate primaries for the super--GZK airshowers.

\subsection{Monopole Acceleration}
\label{sec:accelerate}

The kinetic energy imparted to a magnetic monopole on traversing a magnetic
field along a particular path is \cite{KW96}
\beq
E_K=g \int_{\rm{path}} \vec{B} \cdot\vec{dl}\,
\sim\,g\,B\,\xi\,\sqrt{n}\,
\label{Ekin}
\eeq
where
\beq
g=e/2\alpha=3.3\times10^{-8} \;\rm{esu} \;\;
({\rm or}\; 3.3\times 10^{-8} dynes/G)
\label{charge}
\eeq
is the magnetic charge according to the Dirac quantization condition,
$B$ is the magnetic field strength, $\xi$ specifies the
field's coherence length, and $\sqrt{n}$ is
a factor to approximate the random--walk through the $n$ domains
of coherent fields traversed by the path.
In Table 1 we indicate the cosmic magnetic fields \cite{Kron}
and their
coherence lengths, inferred from observations of synchrotron radiation
and Faraday rotation, and from modeling.
The typical values of the monopole kinetic energies that result
from these fields are also shown in the Table.
A similar table has been constructed in ref.\ \cite{vortons}.

If the early universe dynamics that generated these fields and/or
the present dynamics that maintain these fields were known, then
new Parker--like bounds could be placed upon the extragalactic
monopole flux.  With our limited knowledge at present, efforts
in this direction (latter references in \cite{Parker}) are highly
model--dependent.

The largest energies are seen to come from the magnetic
fields having the longest coherence lengths.
The strength of these sheet fields is inferred from simulations \cite{Ryu97}
and observations \cite{Kim89}.
It is anticipated that in the near future more reliable inferences of the
extra-galactic magnetic fields will become available \cite{Kronberg,Plaga}.
This will allow a firmer prediction of monopole energies.

We emphasize that a typical monopole which travels through the Universe,
and has a mass below the energies indicated in Table 1,
should be relativistic.
Monopoles will gain and lose energy as they random--walk
through the Universe, eventually producing a broad distribution of
energies (with $\Delta E_{K}/E_{K} \, \sim \, 1$)
centered roughly on $\sqrt{n}$ times the typical
energy for a single transit through a region of homogeneous magnetic field.
Here, $n$ is the number of coherent fields encountered in the random walk.
For extragalactic sheets, which we expect to dominate
the spectrum, this number can be roughly estimated to be of order
$n \sim H_0^{-1}/50\,{\rm Mpc} \sim 100$.
The resulting $E_{\rm max}$ is therefore estimated to be $\sim 10^{25}$~eV.
Hence, monopoles with mass below $\sim 10^{14}$~GeV are expected to be
relativistic.  This is a fundamental result.  The rest of this paper is
devoted to the novel phenomenology of relativistic monopoles.
We begin with a discussion of the interactions of monopoles with matter
in the next section, and subsequently calculate monopole signatures
in various detectors.
\begin{table}
\begin{tabular}{|l|c|c|c|r|} \hline
  & $B/\mu$G & $\xi$/Mpc & $gB\xi$/eV & Refs.\\ \hline
normal galaxies & $3 \;{\rm to}\;10$ & $10^{-2}$ & $(0.3\;{\rm
to}\;1)\times 10^{21}$
     &  \cite{Beck96}\\ \hline
starburst galaxies & $10 \; {\rm to}\; 50$ & $10^{-3}$ &
      $(1.7\;{\rm to}\;8)\times 10^{20}$ & \cite{Kronberg}\\ \hline
AGN jets & $\sim 100$ & $10^{-4} \; {\rm to} \; 10^{-2}$ &
            $1.7\times (10^{20}\;{\rm to}\;10^{22})$ & \cite{Kellerman} \\
\hline
galaxy clusters & $5\; {\rm to}\; 30$ & $10^{-4} \; {\rm to} \; 1$ &
     $3\times 10^{18}\;{\rm to}\; 5\times 10^{23}$ & \cite{Ensslin97} \\ \hline
Extragal. sheets & $ 0.1 \; {\rm to} \; 1.0$ & 1 to 30 &
     $1.7\times 10^{22}\;{\rm to}\;5\times 10^{23}$ & \cite{Ryu97} \\ \hline
\end{tabular}
\caption{
Estimated magnetic field strength and coherence length for some
astrophysical environments, and the associated
monopole energies for a single transit through the regions.}
\end{table}

\section{Relativistic Monopole Energy Loss in Matter}
\label{sec:monostop}

GUT monopoles are formed in non-trivial representations of
$SU_{C}(3)\times SU_{L}(2)\times U_{Y}(1)$ and couple to the standard
model gauge fields.  An accurate description of the monopole
stopping power must include all of the standard model
interactions or the relevant subset of those interactions
for the type of monopole considered.  In this paper
we choose to ignore the weak interaction throughout,
which is suppressed in amplitude by factors $\sim M_{Z}^{-2}.$
The electromagnetic and strong interactions will be accounted
for in this section.  

The strong interaction of a monopole is difficult to assess.
Color confinement ensures that all observable monopoles
are color singlet objects,
and so have no classical long--range color--magnetic field.
Nevertheless, we expect $l$--monopoles and baryonic--monopoles to have
very different hadronic interactions. Although $l$--monopoles
are fundamental and lack a color--magnetic charge, the unbroken symmetry in
their core ensures that gluon and light quark
fields will leak out from the center to the confinement
distance $\Lambda_{QCD}^{-1}\;\sim\;1$ fm \cite{E6}.  In this way, l-monopoles
will exhibit some hadronic cross-section in small impact-parameter scattering.
This is probably ignorable.
On the other hand, baryonic--monopoles are intrinsically hadronic in all
partial waves.

We will resume the discussion of the monopole's strong interaction
with hadronic matter after first discussing in some detail
their better--understood electromagnetic interactions.
The electromagnetic interaction of the monopole
may dominate the hadronic interaction
because the electromagnetic coupling of the monopole is large 
\cite{large},
\beq
\alpha_M=\frac{1}{4\alpha}\simeq 34\,,
\label{alphaM}
\eeq
and mediated by a long--range field.
At large distances and high velocities, the
magnetic monopole with Dirac charge mimics the electromagnetic interaction of
a heavy ion of charge
$Z\sim\sqrt{\alpha_M/\alpha}\sim\frac{1}{2\alpha}\simeq 68$.
We will follow others and treat the
electromagnetic interaction of the monopole with matter
semi--classically, i.e., viewing the monopole as a classical source of
radiation, while treating the
matter--radiation interaction quantum--mechanically.
In this way, the large electromagnetic coupling of the monopole
is isolated in the classical field, and the matter--radiation
interaction can be calculated perturbatively \cite{foot2}.

The value of $\alpha_M$ given here, and conservatively used
throughout this paper, is the minimal value due to Dirac.
Other monopole solutions have charges which are integer multiples of
the Dirac charge, and therefore $\alpha_M$'s which are larger by a factor
of $n^2$, $n=1,2,3,\dots$.
For example, the Schwinger solution and the original
t'Hooft-Polyakov monopole have twice the Dirac charge,
and an $\alpha_M$ that is four times minimal.

\subsection{Electromagnetic Interactions}

We consider here the energy losses of the monopole due to the four
electromagnetic processes: ionization resulting from collisions with 
atomic electrons,
$e^+ e^-$ pair production, bremsstrahlung, and the photonuclear interaction.
It will be the pair production and photonuclear interactions,
that dominate the energy loss for very large $\gamma\equiv E/M$ and
will be mostly responsible for the growth of a particle shower.
The monopole-matter electromagnetic interaction for
$\gamma < 100$ is well reported \cite{Giac1,Ahlen1} for
atomic excitations and ionization losses in the
absorber, including the density--dependent suppression effect
arising from charge--polarization of the medium \cite{LL}.
In the literature these two effects are collectively referred
to as ``collisional'' energy losses and we follow that
nomenclature.

It is convenient to express the energy loss per ``length'' in units
of column density (g/$\rm{cm}^{2}$)
which we use throughout this work.
The conversion from true length increment $d\,L$ to column density increment
$d\,x$ is $d\,L=d\,x/\rho_N=(\frac{N_A}{{\rm g}})(\frac{Z}{A\,n_e})\, x$,
the latter form arising from writing the nuclear matter--density $\rho$
as $m_N\,n_N$, and identifying the nucleon mass $m_N$ and density $n_N$
with the first and second fractions, respectively.  In conventional notation,
$N_A$ is Avogadro's number, $Z$ and $A$ are the mean nuclear charge and
number, $n_e$ is the electron density, and g is the gram unit. 
We absorb the mass into $A$, giving $A$ the units of grams per mol.

One approach to electromagnetic interactions is to replace the
monopole current with an equivalent photon flux, 
$dN_{\gamma}/d\omega = 2\alpha_{\rm{M}}/\pi\omega\ln (m\gamma
/\omega ),$ with $m$
being the target mass and $\omega$  the photon energy in the 
rest frame.  The equivalent photon approximation (EPA) is strictly
valid for limited perpendicular momentum transfer
$q_{\perp} \ll m,$ and limited energy transfer
$\omega \ll m \gamma$ \cite{Effphoton}.
Consequently, we prefer to use the more exact monopole--target
scattering formula.  Nevertheless, the EPA presents two illuminating
features, the $1 /\omega$ bremsstrahlung--type of photon
spectrum, and the large normalization $ 2\alpha_{\rm{M}}\pi
=1/2\pi\alpha\approx 22.$  The latter 
tells us that electromagnetic cross--sections are large, while 
the former, in conjunction with $\omega\ll m\gamma,$ 
shows that energy transfers are soft.

\subsubsection{Collisional Energy Loss}

The magnetic monopole stopping power formula
calculated by Ahlen \cite{Ahlen1} includes ionization
of the absorber at small impact parameters and atomic
excitation at large impact parameters.
For highly relativistic monopoles we may ignore the various
correction terms and simply describe ionization with
\beq
\frac{dE_{\rm{coll}}}{dx} = - \frac{4\pi Z
\alpha\alpha_{M} N_{\rm{A}}}{A m_{e}}\left[
\ln \left(\frac{m_{e}\beta^{2} \gamma^{2}}{I} \right)
-\frac{\delta}{2}\right].
\label{eq:monostop}
\eeq
$m_{e}$ is the electron mass,
$I$ is the mean charge and ionization energy of
the material, and $\delta$ parameterizes the 
density effect \cite{Sternheimer}.

\subsubsection{Pair Production}

A thorough study of pair production for muons \cite{Kelner}
is adapted here for the case of a monopole.
To accurately calculate shower development resulting
from pairproduction the inelasticity must be understood.
For a relativistic monopole primary
the energy loss to pair production is
\beq
\frac{dE_{\rm{pair}}}{dx} \simeq - \frac{16}{\pi}
\frac{\alpha^{3}\alpha_{M}Z^{2}N_{\rm{A}}}
{A\,m_{e}}\;\gamma\;\left(\frac{M}{m_{e}}\chi\right),
\label{eq:kelner}
\eeq
where
\beq
\chi\equiv \sum_{i=1}^{5}\chi_{i}\,,
\eeq
\beq
\chi_{i}=\int_{\eta_{\rm{min}}}^{\eta_{\rm{max}}}
d\eta \; F_{i}(\gamma,\eta),
\eeq
the inelasticity, or fraction of monopole energy transferred in the
interaction, is $\eta\equiv \Delta E/E,$ and the five functions 
$F_{i}(\gamma,\eta)$ (given in \cite{Kelner}) are labeled by 
the five distinct regions of inelasticity and nuclear--charge 
screening:
\beq
\ba{lll}
1)\hspace{1cm}&\rm{slow~pairs,~no~screening,}\hspace{1cm}
&\frac{2m_{e}}{\gamma M}<\eta<\frac{2m_{e}}{M} \\
2)\hspace{1cm}&\rm{slow~pairs,~total~screening,}\hspace{1cm}
&\frac{2m_{e}}{\gamma M}<\eta<\frac{2m_{e}}{M} \\
3)\hspace{1cm}&\rm{fast~pairs,~no~screening,}\hspace{1cm}
&\frac{2m_{e}}{M}<\eta<
\frac{\gamma}{\frac{M}{m_{e}}+\gamma} \\
4)\hspace{1cm}&\rm{fast~pairs,~no~screening,}\hspace{1cm}
&\frac{\gamma}{\frac{M}{m_{e}}+\gamma}<\eta<1 \\
5)\hspace{1cm}&\rm{fast~pairs,~total~screening,}\hspace{1cm}
&\frac{2m_{e}}{M}<\eta<1. \\
\ea
\eeq
\begin{figure}[c]
\centerline{\hbox{
\epsfxsize=210pt \epsfbox{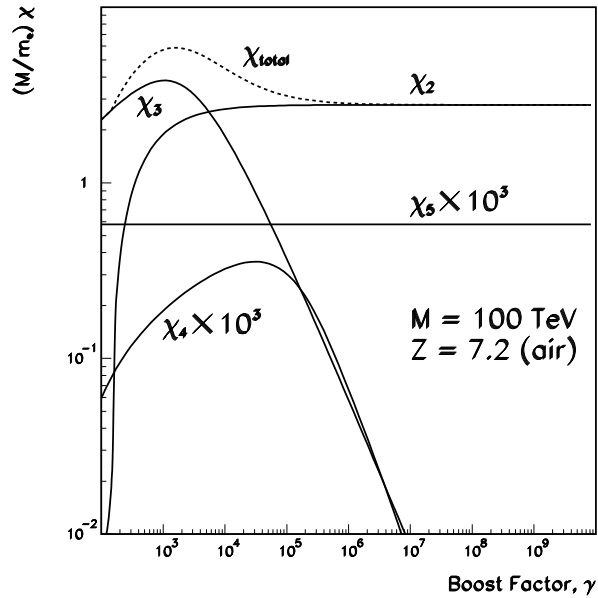}
}}
\caption[Monopole-Induced Pair Production.]
{Electron--pair production of an $M=100$ TeV relativistic
monopole in air is described in terms of the functions $\chi_{i}$
derived in \cite{Kelner}.  The $\chi_{i}$'s relate
to different regions of inelasticity space and either
total or no screening.}
\label{fig:pairint}
\end{figure}

In fig.~(\ref{fig:pairint})
we plot the factors $\frac{M}{m_{e}}\chi_i$ versus $\gamma$ to examine
the relative strengths of the $\chi_{i}$'s.
To a good approximation,
pair production is proportional to the sum of just
$\chi_{2}$ and $\chi_{3}$.
The expression $\chi_{1}$ is unphysical over the
range of $\gamma$ plotted
while $\chi_{4}$ and $\chi_{5}$
have been multiplied by $10^{3}.$
For $M\gg m_{e}$, and
keeping mass corrections with the largest powers of
$\frac{m_{e}}{M}\ln^n\left(\frac{M}{m_{e}}\right)$,
the no screening limit is
\beq
\frac{dE_{\rm{pair}}}{dx} \simeq
-\frac{19\pi}{9} \frac{\alpha^{3}\alpha_{M}Z^{2}N_{\rm{A}}}
{A m_{e}}\;\gamma\left[(1-B_1)
\ln\left(\frac{\gamma}{4}\right)-B_2\right]\,,
\label{eq:pairstop1}
\eeq
and the total screening limit is
\beq
\frac{dE_{\rm{pair}}}{dx} \simeq
-\frac{19\pi}{9} \frac{\alpha^{3}\alpha_{M}Z^{2}N_{\rm{A}}}
{A\,m_{e}}\;\gamma\left[(1-B_1)
\ln\left(\frac{189}{Z^{1/3}}\right)
+\ln 2-B_3\right]\,,
\label{eq:pairstop2}
\eeq
where 
\beq
\ba{l}
B_1=\frac{48}{19\pi^{2}}\frac{m_{e}}{M}
\ln^{2}\left(\frac{M}{m_{e}}\right)\\
B_2=\frac{11}{6}-
\frac{16}{19\pi^{2}}\frac{m_{e}}{M}
\ln^{3}\left(\frac{M}{m_{e}}\right)\\
B_3=\frac{1}{38}+
\frac{16}{19\pi^{2}}\frac{m_{e}}{M}
\ln^{3}\left(\frac{M}{m_{e}}\right).\\
\ea
\eeq
For $E\gg M\gg m_e$ as here,
the contributions of the three $B_i$ to eqs.\ (\ref{eq:pairstop1})
and (\ref{eq:pairstop2}) are negligible.
Examination of fig.~(\ref{fig:pairint}) shows that
for $\gamma\gsim 10^{5}$, a total screening process
dominates, meaning that eq.~(\ref{eq:pairstop2}) 
describes the pair production energy loss.


\subsubsection{Monopole Bremsstrahlung}

Bremsstrahlung radiation by a relativistic particle
in collision is inversely proportional to its mass,
so we expect bremsstrahlung to
be negligible for massive monopoles.
The approximate stopping power due to brehmsstrahlung losses
has been calculated \cite{Jackson} for electric charges
in the negligible nuclear--screening limit \cite{foot3}.
For monopoles it is
\beq
\frac{dE_{\rm{rad}}}{dx} \simeq -\frac{16}{3}
\frac{\alpha\alpha^{2}_{M}Z^{2}N_{\rm{A}}}{A\,M}\,\gamma\,\ln(\gamma).
\label{eq:Ebrem2}
\eeq

\subsubsection{Monopole Photonuclear Interaction}

Virtual photon exchange between a monopole and a nucleus
is described by the photonuclear cross--section.
This process is $MN\rightarrow MX,$ where $M$ is a monopole,
$N$ a nucleus, and $X$ are final state hadrons. 
The energy loss of leptons via the photonuclear interaction
has been re-evaluated recently \cite{Reno} in light of the 
HERA results for real and virtual photon--nucleon scattering. 
A full range of momentum transfer is accounted for 
from real $(Q\rightarrow 0)$ photon exchange to highly
virtual photon--nucleon interactions.  This analysis
is adapted to monopoles with the replacements
$m_{\mu} \rightarrow M$ and $e \rightarrow e/2\alpha$ where
$m_{\mu}, M$ are the muon and monopole masses, respectively.
For $\gamma > 10^{6} $ the photonuclear 
energy loss dominates pair production and
the calculations, based on \cite{Reno}, give roughly
$dE/dx \propto \gamma^{1.28}.$

\subsection{Total Electromagnetic Losses}

We collect the electromagnetic energy loss processes
together and plot them in fig.~(\ref{fig:pairnbrem})
for $M=100$ TeV monopoles.
The dominant interactions are atomic collisions for
$\gamma < 10^{4},$ pair production for 
$10^{4} < \gamma < 10^{6},$ and photonuclear for 
$\gamma > 10^{6}.$
\begin{figure}[b]
\centerline{\hbox{
\epsfxsize=210pt \epsfbox{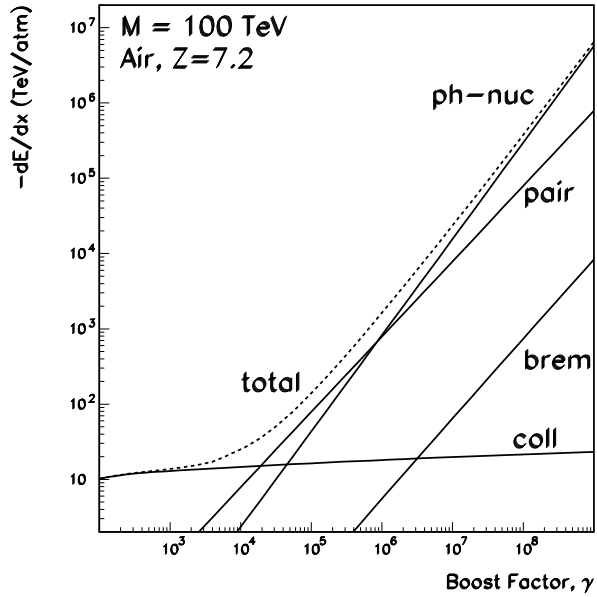}
}}
\caption{The electromagnetic energy loss from collisional, bremsstrahlung,
electron--pair production, and the photonuclear interaction 
of a $100$ TeV relativistic monopole in air.  Collisional, 
pair production, and the photonuclear interaction are roughly
independent of the monopole mass whereas bremsstrahlung is
$\propto M^{-1}.$  The units of energy loss are given in
TeV per atmosphere, where $1\,\rm{atm} = 1033 \,\rm{g}/\rm{cm}^{2}.$}
\label{fig:pairnbrem}
\end{figure}


\subsection{Hadronic Interactions}

GUT monopoles typically contain internal color fields.
The $l$-monopole's color field is soft, extending to a distance
$\Lambda^{-1}\sim 1$~fm (where $\Lambda\equiv\Lambda_{QCD}$).
Its internal color is similar to that of a normal hadron,
however there are no valence quarks internally, and so a better analogy
to the color structure of an $l$-monopole is a glueball.
Thus, we expect the hadronic cross--section of $l$-monopoles to be
of typical hadronic size, with an inelasticity
$\eta \sim \Lambda/M \sim 10^{-10}\,(10^{10}{\rm GeV}/M)$.
For $l$--monopoles an approximate hadronic stopping power would then be

\beq
\frac{dE_{\rm{had}}}{dx}\simeq -\frac{\gamma\,\Lambda}{\lambda}
\simeq - \gamma\,\Lambda\,
\left(\frac{N_{\rm{A}}}{\rm{gram}}\right)
\,\sigma_{\rm{had}}
\label{eq:hadstop}
\eeq
where the mean--free--path between hadronic interactions
is $\lambda = (N_{\rm{A}}\sigma_{\rm{had}})^{-1},$ for
a typical hadronic cross--section $\sigma_{\rm{had}}.$
The magnitude of expression (\ref{eq:hadstop}) appears to be
comparable to the energy loss to electron--pairs discussed above, so
the hadronic interaction for $l$-monopoles may make a
significant contribution to the total energy loss.
Lacking reliable knowledge of the $l$-monopole's hadronic interactions,
we will neglect this possible contribution in the following
signature calculations.  In passing we mention that: 1)
The inclusion of eq.~(\ref{eq:hadstop}) in the longitudinal
shower model (sec.~\ref{sec:monosigns}) would
feed the electromagnetic shower.  For each hadronic interaction
length in the development of a subshower,
about 30\% of the hadronic energy
is transferred to the electromagnetic component; thus,
after a subshower undergoes a few hadronic
interactions, ${\cal{O}}(1)$ of the energy will be in the electromagnetic
shower.  2) The range of mass for which monopole
tomography (to be discussed in in sec.~(\ref{sec:earth}))
is viable will be shifted toward higher mass
with the inclusion of eq.~(\ref{eq:hadstop}).  

The baryonic--monopoles are quite different.
In section (\ref{sec:Mbaryon}) we develop a model for the energy
loss of baryonic--monopoles.
The QCD string model, where the total cross--section
grows with string length $l$, describes baryonic-monopole
hadronic interactions, with the significant caveat
that the confining strings in the monopole are readily stretched
but not easily broken.  The energy loss is estimated to be
\beq
\frac{dE_{\rm{had}}}{dx}(x)\simeq
-\frac{\gamma\,\Lambda}{\lambda(x)}
\simeq - \gamma\,\Lambda\,
\left(\frac{N_{\rm{A}}}{\rm{gram}}\right)
\,\sigma(x)
\simeq - \gamma \,
\left(\frac{N_{\rm{A}}}{\rm{gram}}\right)
\,l(x)\,,
\label{eq:hadstop2}
\eeq
where the string cross--section $\sigma (x)
=l(x)\Lambda^{-1}$ is explicitly a function of
column depth $x$.


\section{Monopole Electromagnetic Signatures}
\label{sec:monosigns}

Signature events for $l$-monopoles are derived below with
a specific emphasis on 1) the general shower development,
2) Cherenkov signatures,
3) the nitrogen fluorescence signature, and
4) Earth tomography.
The general shower characteristics are developed first as
the other signatures are derivable from that model.
For the remainder of this section we
only consider $l$-monopoles, which we will simply refer
to as ``monopoles''.

\subsection{Monopole Shower Development}

Monopoles will be highly
penetrating primaries. On average, there will
be a quasi-steady cloud of secondary particles
continuously regenerated along the monopole 
trajectory.  Thus, we will
call this type of shower ``monopole--induced,''
in contrast to conventional particle--initiated showers.

\subsubsection{Monopole--Induced Subshowers}

Given a fast monopole passing through matter,
the various electromagnetic processes discussed
in sec.~(\ref{sec:monostop}) can inject
energetic photons, electrons, and positrons into the
absorbing medium.
If the energy of these injected secondary particles
is sufficient, they may initiate a particle cascade.
In the simple model we consider, originally developed
by Heitler \cite{Heitler} and reviewed in
\cite{Gaisser,Sokolsky}, the photon pair
production length is set equal to the electron (or positron)
radiation length.  Consider
a photon primary of initial energy $E_{0}$ which
travels a mean distance $R$ through the absorber
before pair producing \cite{foot4}.
The two particles in the produced pair are
assumed to share equally the initial photon's energy.
After traveling another
distance $R$ the electron and positron each
radiate a bremsstrahlung photon where
the produced photon takes half the parent energy.  The 
particle number has doubled again and the energy per 
particle is halved again.  The shower continues
to develop in this geometric fashion
until the energy per particle drops to the
critical energy, $E_{c},$ below which ionization
of the absorber dominates the energy losses due to
particle--production processes.
For typical materials, $E_{c}\simeq 100$ MeV.

From the relation for energy per particle,
\beq
E(x)=E_{0}\,{\rm{e}}^{-\frac{x}{\xi_{e}}}\,,
\eeq
it is apparent that $R$ is related to the radiation
length $\xi_{e}$ by $R=\xi_{e}\ln 2.$
The number of particles in the shower
at a given depth is approximately
\beq
N(x)={\rm{e}}^{\frac{x}{\xi_{e}}}\,.
\eeq
The maximum number produced is
$N_{\rm{max}}=\frac{E_{0}}{E_{c}}$,
from which we infer the depth of the shower maximum to be
\beq
X_{\rm{max}}=\xi_{e}\ln\left(
\frac{E_{0}}{E_{c}}\right).
\label{eq:xmax}
\eeq
After reaching the shower maximum at $X_{\rm{max}}$, the
shower size decreases with column depth as a relativley mild exponential.
The value of $\xi_e$ in air is quoted as $34\,{\rm g/cm^2}$ \cite{Sokolsky},
while the value in ice is quoted as $36\,{\rm g/cm^2}$ \cite{Zas}.
The slight dependence on composition reflects the small differences
in mean charge of the nuclei.
The attenuation length for the post-$X_{\rm{max}}$ decay is 
approximately 200~$\frac{\rm{g}}{\rm{cm}^{2}}$ \cite{Clay}, 
or about $6\,\xi_e$.

\subsubsection{Longitudinal Shower Profile}

It is straightforward to qualitatively describe the electromagnetic 
shower from pair production.
The subshowers begin their existence as an $e^+ e^-$ pair,
produced nearly at rest in the cms.  
The subshower energy is therefore 
$\sim 2 m_e$ boosted to the lab frame.
Since $M$ is much greater than the nucleon mass, the cms frame is 
nearly coincident with the monopole rest frame, and we have
$E_0 \sim 2 \gamma m_e$ for the subshower energy.
The electromagnetic process by which the subshower repeatedly cascades,
and the process by which the monopole initiates the subshower
differ only by the attachment of the monopole to a photon line;
i.e., the monopole interaction mean free path $\lambda$ is a factor 
of $\frac{\alpha_M}{\pi}\frac{1}{\ln 2} \sim 16$ smaller than 
the subshower radiation length $\xi_e$.
Each subshower attains its $N_{\rm max}$ 
and retains this particle number 
for roughly an attenuation length 6~$\xi_e$ (as presented above).
Thus, the number of subshowers contributing to the steady--state
shower size at any given moment is 
$ 6\,\xi_e/\lambda\sim 100$.
With $\sim 100$ subshowers contributing to the total shower,
we find $N \sim 100\,E_0/E_c \sim \gamma$ for the 
total particle number of the steady-state
cloud accompanying the monopole.
It is a pleasant coincidence that the proportionality in
$N \sim \gamma$ is of order unity.

The electromagnetic shower produced by the photonuclear 
process is harder to estimate.  
This process dominates the total 
energy loss for $\gamma > 10^{6}$ and will dominate
the fluorescence yield over that range, but
it's contribution to the electromagnetic shower
is indirect.  
In an Appendix we present a quantitative model 
for the electromagnetic shower from pair 
production and discuss the contribution from the
photonuclear interaction.
In fig.~(\ref{fig:shower}) shown later, we plot the shower size 
$N_{\rm max}$ versus $\gamma$ based on the detailed shower model. 
It is reassuring to see the qualitative agreement
between the detailed calculation and the 
approximate relation $N_{\rm max}\sim \gamma$ with its pleasing 
${\cal O}(1)$~proportionality.   


\subsection{Monopole Cherenkov Signatures}

When a charge travels through a medium with
index of refraction $n,$ at a velocity
$\beta> \frac{1}{n},$ Cherenkov radiation is emitted
\cite{Fermi}.  The Frank-Tamm formula gives the total power 
emitted per unit frequency $\nu$ and per unit length $l$ by a charge
$Ze$,
\beq
\frac{d^{2}W}{d\nu \,dl} = \pi\alpha Z^{2} \nu
\left[ 1 - \frac{1}{\beta^{2}n^{2}} \right].
\label{eq:cherenkov}
\eeq
The maximal emission of the Cherenkov light occurs at an angle
$\theta_{\rm{max}}=\arccos (1/n\beta)$
where $\theta$ is measured from the radiating particle's direction.

\subsubsection{Direct Monopole Cherenkov Emission}

The interaction of a magnetic charge
with bulk matter requires the replacement of factors of $\epsilon$ with
the Maxwell dual factors $\mu$. But $\mu$ and $\epsilon$ are related 
by the index of refraction. The replacement  the electric
charge formulae, eq. (\ref{eq:cherenkov}) with $Z=1$,  
adequate for magnetic monopoles is
$\alpha \rightarrow n^2/4\alpha$, and
leads to an enhancement factor of 4700 for monopoles
interacting in vacuum and 8300 for monopole interactions in water. 
Cherenkov light from an electric charge source is linearly polarized
in the plane containing the path of the source and the direction of
observation.  However, the light polarization 
from a magnetic charge will be rotated 90 degrees
from that of an electric charge which, in principle, offers a 
unique Cherenkov signature for monopoles \cite{Mpolzn}.

The direct Cherenkov signature of a monopole leads to the
best experimental limits at present on the flux of relativistic monopoles.
From the absence of a Cherenkov signal in deep water PMTs the Baikal
Collaboration
reports the limit $F_M\lsim 6\times 10^{-16}{\rm /cm^2/s/sr}$ \cite{Baikal},
while the absent signal in deep ice translates into the AMANDA limit
$F_M\lsim 1.6\times 10^{-16}{\rm /cm^2/s/sr}$ \cite{AMANDA}.
There is a comparable limit of
$F_M\lsim 4\times 10^{-16}{\rm /cm^2/s/sr}$
obtained by the MACRO collaboration
using scintillators and streamer tubes \cite{MACRO}.
Note that these limits are already slightly more restrictive
than the Parker limit.  This work shows that 
the direct Cherenkov signal in the optical should be swamped by
the electromagnetic shower for monopoles with $\gamma > 10^{5}$
(see fig. (\ref{fig:monoshower})).  Such a signal is not seen by
AMANDA \cite{Halzen} and so the AMANDA monopole flux limit
remains the same as that derived from the non-observation of
a direct Cherenkov signal.  
In sec.~(\ref{sec:earth}) we calculate that relativistic 
monopoles can pass through a large portion of the earth 
for $M~\gsim~3$ PeV. 
Given this result, any experimental limit for upgoing 
relativistic monopoles  
should only apply roughly in the mass range $M~\gsim~3$ PeV.

\subsubsection{Coherent Radio--Wavelength Cherenkov Emission}

In addition to the Cherenkov radiation from the bare
monopole charge there is a contribution from the
relativistic $e^{+}$--$e^{-}$ cascade comprising the shower.
Therefore, an estimation of the total power radiated
in Cherenkov light requires the model developed above
for the size of the monopole induced shower.

The approximate lateral width of a monopole-induced
cascade is given by the Moli\`{e}re radius (\ref{eq:moliere})
which means the excess electric charge \cite{foot7}
is confined roughly within a distance $R_{\rm{M}}.$
Cherenkov light of wavelength
$\lambda\gg R_{\rm{M}}$ emitted by the
monopole-induced shower will be radiated coherently
and be strengthened by the $Z^{2}$ factor in
eq.~(\ref{eq:cherenkov}).  Simulations of electromagnetic
showers in ice \cite{Zas} show that the cascade contains an 
electric charge excess of about 20\% the total shower size.
This is plotted for ice in figure (\ref{fig:shower}).

The proposed Radio Ice Cherenkov Experiment (RICE)
may offer the best chance
for the detection of monopoles
in a $1 \rm{km}^{3}$ scale detector
resulting from the transparency of ice in
the radio. \cite{Askaryan}
A thorough study, on a par with that done for
high energy neutrinos \cite{Frichter}, has
not been undertaken for monopoles in RICE.
In lieu of such a study we remark that the most noticeable
signature should be that
the monopole is highly penetrating and will traverse
the full detector size without an appreciable loss of
kinetic energy.  Thus, the monopole
signature is easily distinguished from a neutrino event,
which is a localized shower that produces a Cherenkov cone 
detectable only by a limited number of antennae 
lying within the cone.  Timing can be used to
reconstruct the monopole path.

The analysis of neutrino energy thresholds at RICE 
can be used to estimate an energy threshold for monopole
events.  Figure (13) in \cite{Frichter} can be interpreted 
for {\it{monopole}} detection as showing that an effective 
volume of $1 \rm{km}^{3}$ per antenna
is reached if $dE/dx \gsim dE_{\rm{th}}/dx \approx 3 \times 10^{12}
\,\rm{eV} \rm{cm}^{2}/ \rm{g} \approx 3 \times 10^{3} \,
\rm{TeV}/\rm{atm}.$  This translates into a threshold boost factor,
read off from figure (\ref{fig:pairnbrem}), of
$\gamma \gsim \gamma_{\rm{th}} \approx 10^{6}.$ 
For a RICE monopole search to have an effective volume 
$\gsim 1 \rm{km}^{3}$ the monopole boost is restricted to
$\gamma \gsim 10^{6}.$ 

The monopole flux limits which RICE could set are much
stronger than those currently available
(Baikal, AMANDA, or MACRO \cite{Baikal,AMANDA,MACRO,MACRO2}).
A year of observation in the $ \rm{km}^{3}$
RICE detector without a candidate monopole event
would set a flux limit at least as good as
$ 10^{-18}\;\rm{cm}^{-2}\rm{sec}^{-1}\rm{sr}^{-1},$
three orders of magnitude below the Parker bound
and two orders of magnitude better than the current
bounds.

\subsection{Monopole Fluorescence Signatures}

\label{sec:fluor}

A sensitive probe of high energy air shower development is the nitrogen
fluorescence emitted by the atmosphere.  
The Fly's Eye and HiRes experiments
have fruitfully pioneered this technique.  
The yield in nitrogen fluorescence is $\sim 0.5 \%$ times
the total energy in ionization of the atmosphere.
Since ${\cal{O}}(1)$ of the
energy loss by all processes ends up in ionization of the
atmosphere once the subshowers have ranged out,
we need to evaluate the total stopping power to estimate
the fluorescence yield.
The total energy loss is the sum
of the various processes summarized above:
\beq
\frac{d E_{\rm{total}}}{d x} =
\frac{d E_{\rm{coll}}}{d x} + \frac{d E_{\rm{pair}}}{d x}
+ \frac{d E_{\rm{rad}}}{d x} + \frac{d E_{\gamma-\rm{nuc}}}{d x} 
+ \frac{d E_{\rm{had}}}{d x} .
\eeq
For $l$-monopoles the electromagnetic energy
loss dominates, and in a thin absorber like the
atmosphere, $l$--monopole energy deposition is typically
below threshold \cite{foot11}.
For light $l$-monopoles at extreme $\gamma$, photonuclear
losses dominate and
\beq
\frac{d E_{\rm{total}}}{d x} \approx
- \, 10^{6} \;\frac{1}{\cos\theta_{Z}}\;
\left(\frac{Z}{7.2}\right)^{2}
\left(\frac{14.4}{A}\right)
\left(\frac{\gamma}{10^{8}}\right)^{1.28}\;
 \frac{\rm{TeV}}{\rm{atm}}.
\eeq
On the other hand, baryonic--monopoles \cite{Goldhaber} interact
such that $\frac{d E_{\rm{had}}}{d x}$
dominates the total energy loss.
For baryonic--monopoles
a large energy transfer to the atmosphere is natural.
In section (\ref{sec:Mbaryon}) a model for the baryonic--monopole
interaction is developed.  

\subsection{Earth Tomography with Relativistic Monopoles}
\label{sec:earth}

Direct knowledge about the composition and density
of the Earth's interior is lacking.
Analysis of the seismic data is
currently the best source of information
about the Earth's internal properties \cite{Lay,Geo}.
However, another potential probe would
be the study of highly penetrating particles interacting 
with the Earth's interior.  With such a 
means it may be possible to
directly measure the density profile of the Earth's
interior \cite{foot8}.  Here we show that over a 
range of initial kinetic energies monopoles can pass through the
Earth's interior and emerge with relativistic velocities.  
The results of a numerical calculation, making use of the full
energy loss expressions derived earlier and a
simple model approximating the internal composition of the Earth,
are described below.  Our calculation shows
that relativistic monopoles in the $3\rightarrow 10$ PeV 
(1 PeV $= 10^{15}$ eV) mass range
are ideal for Earth tomography, and that monopoles
may pass through the Earth to initiate up--going cosmic ray events.

\subsubsection{Two Shell Earth Model}

\label{sec:earthmodel}

Dziewonski and Anderson \cite{Anderson} have developed the
preliminary Earth reference model (PREM) which is the
standard Earth model in use today.   The PREM model consists of
eight concentric shells of varying density and
composition.  It is sufficient for the purpose
of demonstrating monopole tomography to
simplify this model to two shells, the mantle and the
core \cite{foot9}.
Both shells are taken to be spherically symmetric.
The core has a radius $R_{\rm{core}}=3.486\times 10^{6}$ m
and a mean mass density $\rho_{\rm{core}}
=11.5 \frac{\rm{g}}{\rm{cm}^{3}}.$
The mantle extends from the core out to the earth's surface
at $R_{\oplus}=6.371\times 10^{6}$ m and has a mean mass
density $\rho_{\rm{mantle}}=4.0\frac{\rm{g}}{\rm{cm}^{3}}.$
We take both shells to be of uniform composition.
The chemical composition of the mantle (in mass)
is approximated by
$\rm{SiO_{2}}$ (45.0\%), $\rm{Al_{2}O_{3}}$ (3.2\%),
$\rm{FeO}$ (15.7\%), $\rm{MgO}$ (32.7\%), and $\rm{CaO}$ (3.4\%),
and that of the core is approximated by
Fe (96.0\%) and Ni (4.0\%) \cite{Kraus}.
From these data we calculate the chemical composition
(in percentage of {\it{molecular type}}) for the mantle as
$\rm{SiO_{2}}$ (40.0\%), $\rm{Al_{2}O_{3}}$ (1.67\%),
$\rm{FeO}$ (11.6\%), $\rm{MgO}$ (43.6\%), and $\rm{CaO}$ (3.23\%),
and for the core as
Fe (96.2\%) and Ni (3.8\%).

The electromagnetic energy loss processes for relativistic
monopoles were analyzed in sec.~(\ref{sec:monostop}).  The
energy loss formulae contain various parameters which need
to be specified for the core and mantle of the Earth.
The density effect includes 
the following dimensionful parameters: \\
the plasma frequency \cite{Ahlen1}
\beq
\omega_{p}\simeq \left\{\ba{ll}
66.2\; \rm{eV}=3.36\times 10^{6}\rm{cm}^{-1} &
\;\rm{(core)} \\
40.4\; \rm{eV}=2.05\times 10^{6}\rm{cm}^{-1} &
\;\rm{(mantle)} \ea \right\},
\eeq
and the mean ionization potential \cite{Ahlen1}
\beq
I\simeq \left\{\ba{ll}
285\; \rm{eV} & \;\rm{(core)} \\
172\; \rm{eV} & \;\rm{(mantle)} \ea \right\};
\eeq
and various numbers in the Sternheimer and Peierls
parameterization \cite{Sternheimer}:
\beq
C\equiv -2\ln \left(\frac{I}{\omega_{p}}\right)-1
\simeq \left\{\ba{ll}
-3.92 & \;\rm{(core)} \\
-3.90 & \;\rm{(mantle)} \ea \right\},
\eeq
\beq
a\equiv -\frac{C+4.606 X_{0}}{(X_{1}-X_{0})^{3}}
\simeq \left\{\ba{ll}
0.137 & \;\rm{(core)} \\
0.136 & \;\rm{(mantle)} \ea \right\},
\eeq
the latter obtained from 
$X_{0}=0.2$ and $X_{1}=3.0$ for both the core
and the mantle.  The functional form \cite{Sternheimer}
for $\delta(\gamma)$ in the core is
\beq
\delta^{\rm{core}}(\gamma)\simeq \left\{\begin{array}{ll}
0 & \;\gamma < 1.87 \\
-3.92+\ln (\gamma^{2}-1)+0.0297
\left[13.82-\ln (\gamma^{2}-1)\right]^{3}
& \; 1.87 < \gamma < 1000 \\
-3.92+ \ln (\gamma^{2}-1) & \; 1000 < \gamma
\end{array} \right\},
\eeq
and in the mantle is
\beq
\delta^{\rm{mantle}}(\gamma)\simeq \left\{\begin{array}{ll}
0 & \;\gamma < 1.87 \\
-3.90+\ln (\gamma^{2}-1)+0.0295
\left[13.82-\ln (\gamma^{2}-1)\right]^{3}
& \; 1.87 < \gamma < 1000 \\
-3.90+ \ln (\gamma^{2}-1) & \; 1000 < \gamma
\end{array} \right\}.
\eeq
Lastly, the mean nuclear charges and 
mean atomic weights, calculated from the 
chemical compositions given above, are
\beq
\bar{Z}\simeq \left\{\ba{ll}
26.1 & \;\rm{(core)} \\
10.8 & \;\rm{(mantle)} \ea \right\},
\eeq
\beq
\bar{A}\simeq \left\{\ba{ll}
56.1~\rm{grams/mol} & \;\rm{(core)} \\
21.7~\rm{grams/mol} & \;\rm{(mantle)} \ea \right\}.
\eeq

\subsubsection{Numerical Results}


We integrate $\frac{d\gamma(x)}{dx}
\equiv \frac{1}{M}\frac{dE(x)}{dx}$ to find the
final boost factor for the monopole after traversing the earth.  
The input parameters are the monopole mass and 
initial boost factor.  The results are plotted in figure
(\ref{fig:mono1}). 
The figure shows several curves
with different initial kinetic energies, for a 
fixed monopole mass.
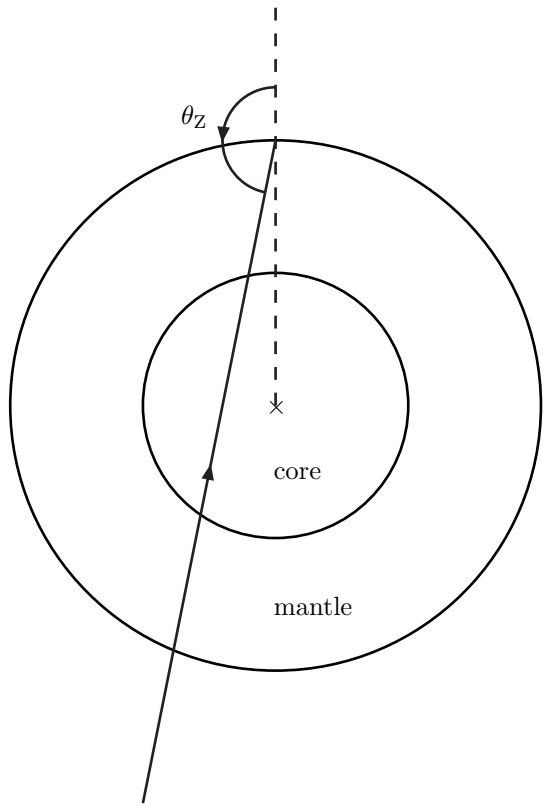
\begin{figure}[t]
\begin{center}
\begin{picture} (250,300)(70,50)
\SetWidth{1.0}
\SetScale{1.}
\GCirc(200,200){100}{1}
\GCirc(200,200){50}{1}
\ArrowLine(150,50)(200,300)
\DashLine(200,200)(200,350){5}
\ArrowArc(200,300)(20,90,258)
\Text(201,200)[c]{$\times$}
\Text(200,175)[l]{core}
\Text(200,125)[l]{mantle}
\Text(175,310)[r]{$\theta_{\rm{Z}}$}
\end{picture}
\end{center}
\caption{A schematic representation
of earth tomography with a highly penetrating particle.
Monopole energy losses differ in passing through
the core or mantle and so affect the energy spectrum of
upcoming monopoles as a function of
zenith angle $\theta_{\rm{Z}}.$  For a zenith angle
$\theta_{\rm{Z}}^{\rm{core}}
\simeq 147^{\circ},$ the upcoming
particle grazes the edge of the core.}
\label{fig:monotom}
\end{figure}
The initial energy spectrum can be determined from
the $\theta_{\rm{Z}}\leq\frac{\pi}{2}$ data.  
These monopoles have not passed through the earth
and therefore retain the original kinetic energy
attained from the large-scale magnetic fields.
A comparison of these data
with the $\theta_{\rm{Z}} > \frac{\pi}{2}$ data could
make a determination of the earth's interior possible. 
In figure \ref{fig:mono1}, the discontinuity due to 
the mantle--core boundary is clear \cite{foot10}. 
The zenith angle for a trajectory tangent to the
core boundary is
\beq
\theta_{\rm{Z}}^{\rm{core}}=\pi
-\arcsin\left(\frac{R_{\rm{core}}}{R_{\oplus}}\right)
\simeq 147^{\circ}.
\eeq

Assuming that ${\cal{O}}(100)$ data points would 
be sufficient to delineate the attenuation curves
of figure \ref{fig:mono1}  
we can estimate the detector exposure needed in the 
$2\pi$ steradians projected through the Earth:
\beq
t\,=\,1\,\rm{year}\;\frac{1}{2\pi}\left(\frac{N}{100}\right)
\left(\frac{10^{-16}/\rm{cm}^{2}\,\rm{sec}\,\rm{sr}}{F_{\rm{M}}}\right)
\left(\frac{1\rm{km}}{R}\right)^{2}.
\eeq
The monopole flux $F_{\rm{M}}\approx
10^{-16}/\rm{cm}^{2}\,\rm{sec}\,\rm{sr}$
is roughly the experimental upper bound 
of \cite{Baikal,AMANDA,MACRO}.

\begin{figure}[c]
\centerline{\hbox{
\epsfxsize=210pt \epsfbox{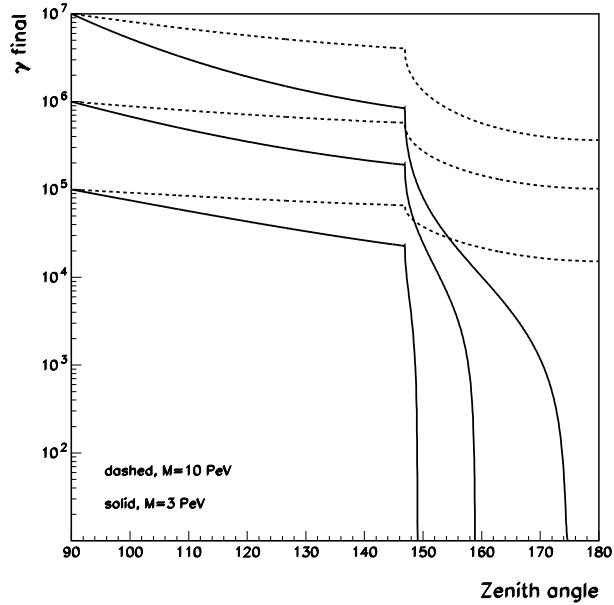}
}}
\caption{$M=3\times 10^{15}$ and $M=10^{16}$ eV monopoles
passing through the earth and emerging with a diminished boost 
factor for increased zenith angle.  
The range of initial kinetic energies presented here is
$3\times 10^{20}\rm{eV}<E_M< 10^{23}$ eV.}
\label{fig:mono1}
\end{figure}

\section{Baryonic--Monopole Air Showers And Super--GZK Events}
\label{sec:Mbaryon}

The observation of air--showers above the
GZK cutoff at $\egzk\sim 5\times 10^{19}$~eV presents a puzzle.
These events cannot be due to nucleons or photons propagating
from sources located at cosmic distances (see \cite{KW96}
for a recent discussion of the puzzling nature of these events).
The natural acceleration of monopoles to energies above the
GZK cutoff,
and the allowed abundance of a monopole flux at the observed super--GZK
event rate of $2.0\pm0.5\times 10^{-20}/{\rm cm}^2/{\rm s}/{\rm sr}$
above $10^{20}$~eV \cite{ringberg} (five orders of magnitude below the
Parker limit), 
motivates us to ask whether monopoles may contribute to the
observed super--GZK events.
As a proof of principle, here we present a simple
model of a baryonic--monopole interaction in air which produces
a shower similar to that arising from a proton primary.
(We note that an unrelated mechanism for mimicking a high--energy 
proton--initiated air shower with a monopole--nucleon bound 
state has been advanced in \cite{vortons}.)
To mimic a proton--induced shower the monopole must transfer
nearly all of its energy to the shower in a very small distance.
The large inertia of a massive monopole makes this impossible if
the strong cross--section is typical, $\sim 100$~mb \cite{Minertia}.
The cross--section we seek needs to be much larger.
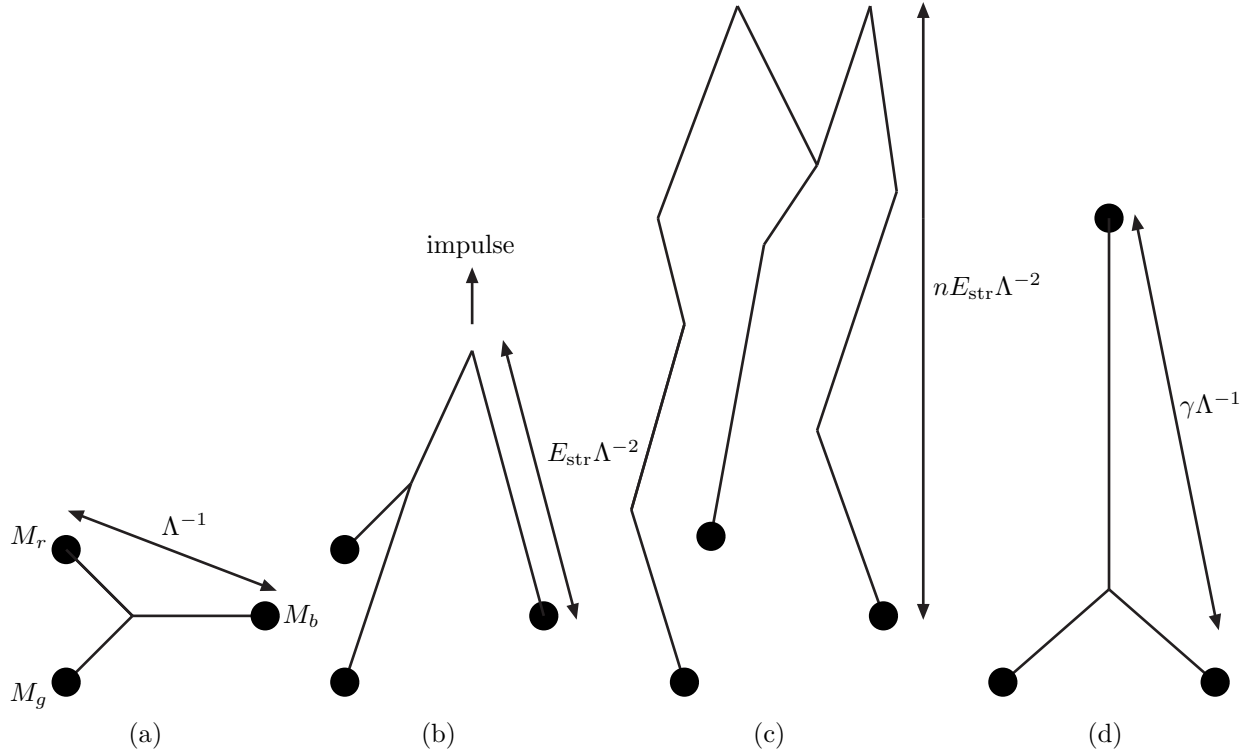
\begin{figure}[b]
\begin{center}
\begin{picture} (450,300)(0,0)
\SetWidth{1.0}
\SetScale{1.}
\Line(32,50)(82,50)
\GCirc(82,50){5}{0}
\Line(32,50)(7,25)
\GCirc(7,25){5}{0}
\Line(32,50)(7,75)
\GCirc(7,75){5}{0}
\Text(0,20)[r]{$M_g$}
\Text(0,80)[r]{$M_r$}
\Text(89,50)[l]{$M_b$}
\LongArrow(12,88)(85,60)
\LongArrow(12,88)(9,89.15)
\Text(52,85)[c]{$\Lambda^{-1}$}
\Text(37,5)[c]{(a)}
\Line(137,100)(112,25)
\GCirc(187,50){5}{0}
\Line(137,100)(112,75)
\GCirc(112,25){5}{0}
\Line(137,100)(160,150)
\GCirc(112,75){5}{0}
\Line(160,150)(187,50)
\LongArrow(174,145)(199,50)
\LongArrow(174,145)(172,152.6)
\Text(189,112)[l]{$E_{\rm{str}}\Lambda^{-2}$}
\LongArrow(160,160)(160,180)
\Text(160,190)[c]{impulse}
\Text(148,5)[c]{(b)}
\Line(240,25)(220,90)
\Line(220,90)(240,160)
\Line(240,160)(230,200)
\Line(220,90)(240,160)
\GCirc(240,25){5}{0}
\Line(250,80)(270,190)
\GCirc(250,80){5}{0}
\Line(315,50)(290,120)
\Line(290,120)(320,210)
\Line(320,210)(310,280)
\GCirc(315,50){5}{0}
\Line(230,200)(260,280)
\Line(260,280)(290,220)
\Line(270,190)(290,220)
\Line(290,220)(310,280)
\LongArrow(330,200)(330,50)
\LongArrow(330,200)(330,280)
\Text(335,175)[l]{$nE_{\rm{str}}\Lambda^{-2}$}
\Text(273,5)[c]{(c)}
\Line(440,25)(400,60)
\GCirc(440,25){5}{0}
\Line(360,25)(400,60)
\GCirc(360,25){5}{0}
\Line(32,50)(7,75)
\GCirc(400,200){5}{0}
\Line(400,60)(400,200)
\LongArrow(440,50)(410,200)
\LongArrow(440,50)(441,46)
\Text(440,130)[c]{$\gamma\Lambda^{-1}$}
\Text(400,5)[c]{(d)}
\end{picture}
\end{center}
\caption{A schematic representation of a baryonic--monopole
interacting with the atmosphere that depicts the effect
of string--nucleon interactions.  Fig.~(a) shows the 
baryonic--monopole in its unstretched state, before hitting
the atmosphere.  After the first string--nucleon interaction
(fig.~(b)) the string stretches to length
$\sim~E_{\rm{str}}~\Lambda^{-2}$,
and after $n$ interactions (fig.~(c)) it stretches to
$\sim~n~E_{\rm{str}}~\Lambda^{-2}$ .  Fig.~(d) shows string
stretching from $q$--monopole recoil.}
\label{fig:puffdaddy}
\end{figure}

We model our arguments on \cite{Goldhaber}
where three $q-$monopoles are confined
by $Z_{3}$ strings of color--magnetic flux to form a color--singlet
baryonic--monopole as in figure (\ref{fig:puffdaddy}a).
All scales in the ground state baryonic-monopole are set by the QCD 
scale $\Lambda$.  In particular, the string tension has the usual
QCD-strength $\mu\simeq\Lambda^2$.
We further assume that:\\
(1) Before hitting the atmosphere,
the baryonic--monopole's cross--section is roughly hadronic,
$\sigma_{0} \sim \Lambda^{-2}$.\\
(2) Each interaction between the baryonic-monopole and an air nucleus
transfers on average energy $E_{\rm str}$ (in the monopole rest-frame) into 
the chromomagnetic string system of the baryonic--monopole.\\
(3) Since the chromomagnetic strings can only be
broken with the formation of a heavy $q-$monopole--antimonopole pair, 
the internal energy does not break the strings but rather stretches 
them by amount
$\delta L=E_{\rm str}/\mu\simeq(E_{\rm str}/\Lambda)$~fm;
the string system grows linearly with each interaction, until other emission 
processes of the stretched (and presumably vibrating) string dominate
\cite{foot12}.\\
(4) The cross--section for the interaction of the baryonic--monopole
with a nucleus is geometric,
$\sigma\sim L\Lambda^{-1}$, where $L$ is the (growing) string length;
this assumption says that string--nucleon interactions dominate over 
$q$--nucleon interactions.\\
(5) The energy emission rate from the excited string system 
(in the monopole frame) is proportional 
to some small power $p$  
of the (growing) string length
(and therefore, to the same power of the number of hits and 
(growing) cross-section).

Together, the assumptions imply that the baryonic--monopole 
cross-section grows 
from typical hadronic at first interaction, 
$\sigma_0\sim\Lambda^{-2}$, to 
\beq
\sigma_n=L\Lambda^{-1}=n\,\frac{E_{\rm str}}{\Lambda^3}
\label{sigman}
\eeq
after $n$ interactions.
A scematic drawing of the baryonic--monopole is given in 
figure (\ref{fig:puffdaddy}) for (a) the ground state, 
(b) after the first interaction, 
and (c) after the $n^{\rm th}$ interaction.
This growth ($\propto L$) continues with each interaction, 
until finally the rising rate of particle emission 
$\dot{E}_{\rm emit}$  ($\propto L^p$)
equals the rate of energy acquisition due to stretching.
If $E_{\rm str}\gg \Lambda$, then 
already after the {\sl first} interaction the cross--section is
sufficiently large to shrink the subsequent interaction length
to a small fraction of the depth of first interaction.  
This allows ${\cal{O}}(1)$ of the incident monopole energy to be 
transferred to the air nuclei over a short distance,
just as in a hadron--initiated shower.

We can be more quantitative.
The rate of internal energy acquisition effecting string--stretching 
exceeds the rising rate of particle emission as long as 
\beq
\frac{E_{\rm str}}{\Delta t} > \dot{E}_{\rm emit}\,,
\label{rates}
\eeq
where $\Delta t$ is the time between interactions 
in the monopole rest frame.
Explicitly, we have 
\beq
(\Delta t)^{-1}=\gamma\,(\Delta t_{\rm lab})^{-1} 
=\gamma\frac{c}{\lambda_{\rm lab}} =c\gamma\rho_N \sigma\,,
\label{Deltat}
\eeq
where $\rho_N$ is the nucleon density of air in the lab frame.
It is important to the model that the monopole first has a growing phase,
and then before it becomes non-relativistic, the emission phase.
This requires that inequality (\ref{rates}) hold for the initial interactions.
Such will be the case when $p>1$, i.e. the emission rate from the string
grows nonlinearly with the string length.  We comment on this below.
Inputting eqs.\ (\ref{sigman}) and (\ref{Deltat}) into (\ref{rates}),
we find that the baryonic--monopole remains in its growth phase
until the critical $n^{\rm th}$ interaction
\beq
n_{\rm str}=\frac{\Lambda^2}{E_{\rm str}}
\left[\frac{E_{\rm str}}{\Lambda}
\frac{c\gamma\rho_N}{\kappa(p)}\right]^{\frac{1}{p-1}}\,.
\label{criticaln}
\eeq
Here, $\kappa(p)$ is the proportionality constant in
$\dot{E}_{\rm emit}=\kappa(p) L^p$.
The ``size'' of the monopole at the conclusion of its growth phase is
$\sigma_{n_{\rm str}}=n_{\rm str}E_{\rm str}/\Lambda^3$.
(For our order of magnitude estimates, we  
benignly neglect the recoil of the monopole and 
the related decrease of $\gamma$ with each interaction.)
The distance between the first interaction and the onset of 
the emission phase is 
\beq
\Delta X \sim \sum_{n=1}^{n_{\rm str}}\lambda_{n}\, ,
\label{dx1}
\eeq
which, according to $\lambda_{n}=\Lambda\lambda_0/E_{\rm{str}}n$, is
\beq
\Delta X \sim \frac{\Lambda}{E_{\rm str}}\,\lambda_0
\sum_{n=1}^{n_{\rm str}}\frac{1}{n}\,;
\label{dx2}
\eeq
$\lambda_0$ and $\lambda_n$ are the mfp's of the monopole
in the initial ground state and after $n$ interactions, respectively,
and the density of nucleons is taken here to be a constant
for simplicity of illustration.
The sum is a finite series, and for $n_{\rm str}\gg 1$ is very nearly
equal to $\ln (n_{\rm str})$ \cite{GR}, giving
\beq
\Delta X_{\rm str} \sim 
\frac{\Lambda}{E_{\rm str}}\,\lambda_0\,\ln (n_{\rm str})
\label{dx3}
\eeq
as the estimate for the total distance traveled 
by the baryonic--monopole in its growth phase.
As derived here, $\Delta X$ and $\lambda$ are lengths in the 
monopole rest-frame.  However, our eqs.\ (\ref{dx1})--(\ref{dx3})
are homogeneous in these variables, and so also apply 
in the Earth frame without change.
In terms of column density of air, this distance is just
\beq
\Delta x_{\rm str} \,=\,\frac{n_N}{N_A}\,\Delta X\, 
\,\sim 100\,\left(\frac{\Lambda}{E_{\rm str}}\right)\,
\ln (n_{\rm str})\;{\rm g/cm}^2\,,
\eeq
where the numerical value is evaluated in the Earth frame.
Turning to the emission phase,
equilibrium between absorption and emission fixes the  average 
energy emitted per interaction to be $\sim E_{\rm str}$ in the 
monopole frame.
An energy $M$ in the monopole frame, isotropically emitted,
boosts to $\gamma M=E$ in the lab frame.
Therefore, the number of interations required
to absorb and re-emit the energy observed as an air-shower is
\beq
n_{\rm emit}\sim \frac{M}{E_{\rm str}}\,.
\label{nemit}
\eeq
Numbers are Lorentz invariants.
Using 
$\lambda_{n_{\rm str}}=\lambda_1/n_{\rm str}=
\frac{\Lambda}{E_{\rm str}} \frac{\lambda_0}{n_{\rm str}}$
and eq.\ (\ref{dx3}),
the distance traveled in the emission phase is then 
\beq
\Delta X_{\rm emit}\,\sim\,n_{\rm emit}\,\lambda_{n_{\rm str}}
=\frac{M\Lambda}{E^2_{\rm str}}\,
\frac{\lambda_0}{n_{\rm str}}\,.
\label{dxemit}
\eeq
This too can be quite short compared to the mfp $\lambda_0$
of the ground-state monopole.

The two constraints on the model, that the length of 
the stretching phase  
$\Delta X_{\rm str}$ and 
the length of the emission phase $\Delta X_{\rm emit}$ 
be short compared
to the mfp $\lambda_0$ for first interaction,
can be written with the help of eqs. (\ref{dx3}) and (\ref{dxemit})
as:
\beq
\frac{M\Lambda}{E^2_{\rm str}} \ll n_{\rm str} \ll
e^{E_{\rm str}/\Lambda}\,,
\label{constraints}
\eeq
with $n_{\rm str}$ given by eq.\ (\ref{criticaln}).
For a monopole mass of 100~TeV ($10^{10}$~GeV),
the left and right terms of the inequality are correctly ordered
for $E_{\rm str}/\Lambda > 10$~(20).
Thus, for $E_{\rm str}\gg \Lambda$, 
a broad range of $n_{\rm str}$ values
will satisfy eq.\ (\ref{constraints}).
Furthermore, all values of 
$n_{\rm str}$ are available by suitably choosing the 
free parameter $\kappa (p)$. 
Thus, the model is natural in that it 
does not require fine--tuning of parameters.
We have succeeded in constructing a 
stretchable chromomagnetic--string model for 
the baryonic--monopole which 
provides an existence proof that a very massive monopole,
despite its inertia, may nevertheless 
transfer ${\cal{O}}(1)$ of its relativistic energy to an air shower over a
very short distance.  
Such a monopole mimics the signature of a primary nucleon
or nucleus.  Of course, the last gasp of the monopole will not resemble
a nucleon, for the monopole will eventually become non-relativistic
and shed its internal energy more and more isotropically.
If this ``end game'' energy is a small fraction of the air-shower
energy, it will be difficult to observe.

Let us discuss assumptions (2), (3), and (5).
Any assumption that does not violate known physics is valid
until disproven by Nature.
However, it is motivational to look for plausibility arguments.
Assumption (2) states that energy $E_{\rm str}$ 
per interaction stretches the string.
An analogy to $p-p$ scattering may be illuminating here.
The baryonic monopole contains a QCD energy of roughly $\Lambda$.
Thus, the stong--interaction scattering dynamics bear some 
resemblance to $p-p$ scattering at 
$s\sim \gamma\Lambda m_N\sim ({\rm TeV})^2$.  
$p-p$ scattering data at 2~TeV cms energy 
are available from Fermilab's $p-{\overline p}$~Tevatron collider.
A mainly diffractive QCD interaction stretches particle production 
throughout the rapidity plateau, emitting $\sim 10^2$ particles 
with typically $\sim$~few~GeV energies.
When a $Z_{3}$ string is struck by an air nucleus, 
the diffractive QCD dynamics may be similar,
stretching the colored strings across the rapidity plateau.
However, by assumption (3), the chromomagnetic string cannot break,
and so a significant fraction of 
the $\cal{O}$(100~GeV) energy remains in
the stretched string, with the remaining energy 
materializing as light hadrons.
The value $E_{\rm str}\sim \cal{O}$(100~GeV), implying 
$E_{\rm str}/\Lambda \sim 10^3$, appears more than ample 
(by an order of magnitude or more) to sustain the model
we have outlined.
We also note that whereas the color-electric charge
interactions are asymptotically free, 
the strength of the color-magnetic interactions 
increases with energy to preserve the Dirac quantization condition. 
This may further argue in favor of the  
non-breaking, coherent behavior of the chromomagnetic string.

Assumption (3) states that energy emission by the excited string
is negligible in the early growth phase of the string.
Some analogies may motivate this assumption.
First of all, as is the case with a classical violin string, 
the Q of a resonating string (related to the decay time) 
can be very large compared to the time it takes to excite
(``pluck'') the string.
Secondly, as is the case with cosmic strings, energy loss
through particle emission may be inefficient until cusps 
or interconnections are formed.  
%
Assumption (5) states that the particle emission rate 
from the string grows nonlinearly with increasing $L$.
We believe this is a reasonable expectation upon nature.
E.g., the emission may occur as a result of string pinching
to form cusps, or crossing to form reconnections, etc.
The nonlinear nature of pinching and crossing is the 
reason these phenomena are studied numerically on a lattice,
rather than analytically.
Also, the QCD string is non-abelian, so that interactions 
among string sections are not linear.
In fact, one may wonder whether our assumption (3), namely, 
{\sl linear} string growth per interaction, is sensible.
What really drives the model is that the 
baryonic--monopole cross--section
grows rapidly upon interaction,
and that the particle emission rate grows even more rapidly;
linearity of the string growth, 
adopted in the model here for illustration, 
is not an essential aspect of the dynamics.

According to eq.\ (\ref{sigman}), the geometrical cross--section
after the first interaction grows to a very large value:
\beq
\sigma_{1}\sim
\left( \frac{E_{\rm str}}{100{\rm GeV}}\right)\times 10^4\;{\rm mb}\,.
\eeq
Inspired by such a large cross--section,
one may ask how many interactions occur within 
the first--struck nucleus.
Naively, the answer appears to be extremely large.
The number of interactions is given by counting mfp's
as the monopole travels through the nucleus:
\beq
N_{\rm int}\sim \int^{A^{1/3}\,{\rm fm}}_0 
dz\,[\lambda^{-1}=n_N\,\sigma]\,,
\label{Nint}
\eeq
where here the appropriate nucleon density is 
that in a nucleus, $n_N\sim{\rm fm}^{-3}$.
Note that $N_{\rm int}$ is a Lorentz invariant,
as $z$ and $\lambda$ each scale the same way under a boost.
For the cross-section above, one gets 
$N_{\rm int}\sim E_{\rm str}/\Lambda$.
However, only that part of the string within the geometric 
cross section of the nucleus interacts.  Furthermore, 
there is a causal limit on the rate at which the
chromomagnetic string expands, and therefore on the 
rate at which the monopole cross-section grows.
For a relativistic monopole, one expects the struck string
to expand at nearly the speed of light.  This leads to 
a much more sober estimate for the number of interactions per
first-struck nucleus, $N_{\rm int}\sim A$.
The number of interactions per subsequent struck nucleus is 
similar, as the cross-section saturates at the nuclear size
$A^{2/3}{\rm fm}^2$.
On the other hand, the number of struck nuclei quickly becomes enormous.
With the monopole traveling at near light speed, and the geometric 
cross-section growing also at near $c$, 
essentially all of the nuclei in the forward light--cone starting
at the site of first interaction are struck, until the 
baryonic--monopole is quickly brought to rest.

In passing, we note an alternative possibility for 
a growing strong cross--section might be to
enhance ($q$--monopole)--nucleon hard scattering relative to
string--nucleon scattering, 
so that $q$--monopole recoil in the first interaction 
provides a large stretch to the string. 
The recoil energy can be significant, up to $\gamma m_N/2$,
providing a stretch $\delta L\sim \gamma\Lambda^{-1}$,
when the condition $M^2 \lsim 2E_Mm_N \sim ({\rm PeV})^2$ holds. 
A schematic of $q$--monopole recoil is shown in figure (5d).
The simplest way to increase the ($q$--monopole)--hadron 
cross--section 
is to increase the color-magnetic charge of one or more 
$q$--monopole constituents of the baryonic monopole. 
Increasing one constituent charge 
by an order of magnitude increases the cross-section 
by two orders of magnitude. 
Generally, the values of the constituent charges are model-dependent, 
with larger values typically coming from less attractive models.
An added complication with a large 
($q$--monopole)--nucleon cross--section
is that the energy losses of baryonic--monopoles propagating 
through cosmic media 
and through the magnetic fields of the earth, sun,
galaxy, etc. may require a  re-analysis.

The baryonic--monopole's mfp $(\sigma_0\rho_{N})^{-1}$ for 
the first hadronic interaction is an observable
and therefore of interest.  
We may estimate the unstretched baryonic--monopole ``size'' 
$\sigma_{0}$ by equating  
the energy stored in the string with that due to the 
repulsive magnetic force between the $q$--monopoles.
This gives 
\beq
\mu L_0 \simeq \frac{g^{2}}{L_0}\;.
\eeq
Thus, the unstretched monopole's string length is
\beq
L_0\sim \frac{g}{\sqrt{\mu}}\simeq \frac{6}{\Lambda}\;,
\;{\rm{where}}\; g= \sqrt{\frac{137}{4}}\simeq 6.
\eeq
The lateral size of the string is $\Lambda^{-1}$ so
that the geometric cross section is
\beq
\sigma_{0} \sim 6\Lambda^{-2}
\sim 60\; \rm{mb} \left(\frac{200 \rm{MeV}}{\Lambda}\right)^{2}\;.
\eeq
This is
somewhat larger than our order of magnitude estimate
$\Lambda^{-2},$ but comparable to the cross--section
of a high energy proton.  Accordingly, we expect $X_{\rm{max}}$
for the baryonic--monopole to be similar to that of a
proton, or even a nucleus.

We remark on a possible 
signature discriminating the monopole from a proton.
The large number $n_{\rm emit}\sim M/E_{\rm str}$ 
of soft individual nuclear 
interactions underlying the monopole--induced shower event, 
in contrast to one of a few hard interactions from a primary proton,
may lead to a larger hadron to electromagnetic ratio
in the shower; this in turn leads to a higher muon content.

Finally, we remind the reader that the results of this section 
are predicated on a very speculative model wherein the 
strong cross--section of the baryonic--monopole grows very rapidly
after the first interaction,
and the particle emission rate from excited strings grows even
more rapidly.  While these dynamics are plausible, Nature may reject them.
Then baryonic--monopoles will look much like $l$--monopoles in 
their interactions with matter.

\section{Summary and Conclusions}
 \label{sec:concl}

The challenge to seek a cosmic magnetic monopole flux below the 
Parker limit
has now been met technologically in various underground/ice experiments.  
Here we provide theoretical motivation for further
efforts to increase the search sensitivity.

The Kibble mechanism predicts possibly observable fluxes for monopoles
with masses in the intermediate range, $10^{8\pm 3}$~GeV.
Such monopoles will be naturally accelerated by
cosmic magnetic fields to relativistic energies.
The Universe is transparent to relativistic monopoles meaning
that a cosmic monopole flux should arrive at earth unattenuated
and enter terrestrial detectors.

The possible signatures by which a
relativistic monopole flux may be identified are varied.
The Dirac condition for the monopole electromagnetic
charge is $\alpha_M=1/4\alpha$.
This large coupling makes promising a search for
electromagnetic signatures including direct Cherenkov emission, and
coherent radio--Cherenkov emission from the charged $e^-$--$e^+$ shower.
The radio signal may be observable for monopoles in 
polar ice.  Of interest is the possibility of ``m--raying'' the 
interior of the earth with the monopole flux if their mass is
within the range $3\rightarrow 10$~PeV.

The hadronic properties of monopoles are not well known.
Monopoles with color ($q$-monopoles) and without color ($l$-monopoles)
are expected to be produced in phase 
transitions which break the gauge symmetry to the Standard Model.
Just as the monopole magnetic charge is dual to electric charge,
the monopole chromomagnetic charge is dual to the familiar chromoelectric
charge.
The $l$-monopoles carry QCD fields in their unbroken core,
and so have a soft strong--interaction of size
$\sigma\sim\Lambda_{QCD}^{-2}$.
The $q$-monopoles may have a more interesting strong--interaction.
Presumably they are confined by chromomagnetic strings
into color--singlet baryonic--monopoles, with initial cross-section
$\sigma_0\sim\Lambda_{QCD}^{-2}$.
In the process of interacting, the strings may not break until
stretched to length at least $L\sim M/\mu$,
where $\mu\sim \Lambda_{QCD}^{2}$ is the string tension.
Thus, this stretching (if it can take place) will lead to a much larger cross--section
if the energy-loss mechanisms of the string are slow compared 
to the interaction time.
We have speculated that this enhanced cross--section may be many times larger than
the usual QCD cross--section, and if so it could very quickly transfers ${\cal{O}}(1)$
of the monopole energy to the shower.  In this case baryonic--monopoles would be viable
candidates for the observed cosmic rays above $E_{GZK}\sim 5\times10^{19}$~eV.

\section{Appendix: shower development model}

A monopole is highly penetrating and, as such, can
initiate many subshowers before stopping.  However,
for a subshower to develop, the energy
injected into the absorber in any single interaction
must be greater than $E_{c}$.
Thus, the inelasticity restriction 
$\eta \gsim\frac{E_{c}}{E_{0}}\simeq 10^{-12}
\left(\frac{E_{0}}{10^{20}\rm{eV}}\right)^{-1}$
holds for monopole-matter interactions
which initiate subshowers,
with lower inelasticity events contributing 
only to ionization.

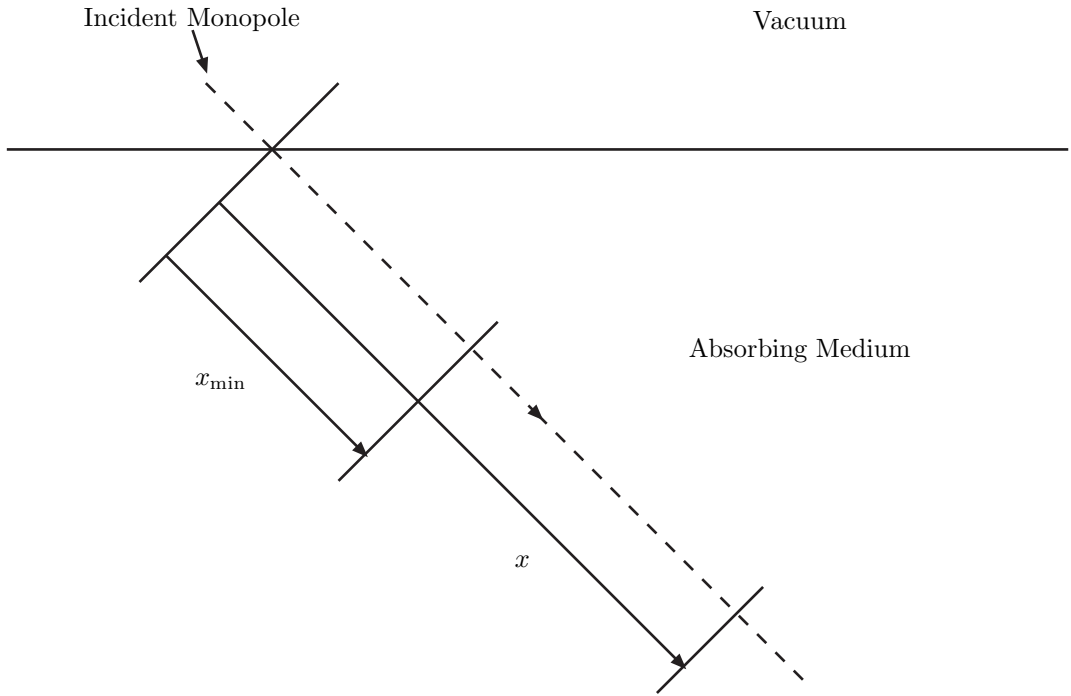
\begin{figure}
\begin{center}
\begin{picture} (400,300)(0,40)
\SetWidth{1.0}
\SetScale{1.}
\Line(0,250)(400,250)
\DashArrowLine(100,250)(300,50){5}
\DashLine(75,275)(100,250){5}
\Line(50,200)(125,275)
\Line(245,45)(285,85)
\Line(125,125)(185,185)
\LongArrow(80,230)(255,55)
\LongArrow(60,210)(135,135)
\LongArrow(70,295)(75,280)
\Text(70,300)[c]{Incident Monopole}
\Text(300,300)[c]{Vacuum}
\Text(300,175)[c]{Absorbing Medium}
\Text(195,95)[c]{$x$}
\Text(81,163)[c]{$x_{\rm{min}}$}
\end{picture}
\end{center}
\caption[A Monopole-Initiated Shower.]{A schematic
representation of the variables entering into a
calculation of the shower development for a
monopole-induced shower.  The monopole path
is along the dashed line.
The depth where the
shower size is evaluated is at $x.$
The smallest depth where
an initiated subshower contributes to the total
shower is $x_{\rm{min}}.$}
\label{fig:monoshower}
\end{figure}

Suppose the monopole initiates
the $j^{\rm{th}}$ subshower at a depth $x_{j}.$  The
subsequent development of the $j^{\rm{th}}$-subshower has
particle number
\beq
N_{j}(x)=N_{0}\;\rm{e}^{\frac{(x-x_{j})}{\xi_{e}}}
\eeq
where $0 < (x-x_{j}) < \xi_{e}\ln\left(\frac{E_{j}}
{N_{0}E_{c}} \right),$ the total
energy injected into the shower at the point $x_{j}$
is $E_{j},$ and $N_{0}$ is the initial
number of particles injected into the shower at
the point $x_{j}$  (For electron pair production
$N_{0}=2,$
whereas for bremsstrahlung and monopole-electron
elastic scattering $N_{0}=1.$).

The total shower development $N(x)$ is then obtained by
summing over all the subshowers
\beq
N(x)=\sum_{j}N_{j}(x).
\label{eq:showersum}
\eeq
The approximate distance between the initiation of subshowers
is given by the monopole mean free path, $\lambda=\frac{1}
{\sigma n},$ where $\sigma$ is the monopole cross-section 
and $n$ is the number density of
scattering centers.  The $j^{\rm{th}}$ interaction then
will roughly occur at the depth $x_{j}=j\lambda.$  If
the inelasticity per interaction is
approximately given by a constant value of
$\eta$ then the energy injected
at the $j^{\rm{th}}$ scattering is
\beq
E_{j}=\frac{\eta}{1-\eta}(1-\eta)^{j}E_{0}\,.
\label{eq:Ej}
\eeq
The monopole will continue to initiate subshowers until
\cite{foot13}
it is degraded in energy to the point where
$E_{j}\simeq N_{0}E_{c}.$  That fixes the maximum number
of subshowers to be
\beq
j_{\rm{max}} =\frac{\ln\left(\frac{(1-\eta)N_{0}E_{c}}
{\eta E_{0}}\right)}{\ln(1-\eta)}.
\eeq
Note that this relation is just a restatement of the condition
$0 < (x-x_{j}) < \xi_{e}\ln\left(\frac{E_{j}}
{N_{0}E_{c}} \right)$.
The sum in eq.~(\ref{eq:showersum}) can now be performed
from $j=1$ to $j=j_{\rm{max}}$.

The sum in eq.~(\ref{eq:showersum}) can be transformed
into an integral over column depth by making a
continuum approximation.  The substitutions
$\sum_{j}\rightarrow\int\frac{dx_{j}}{\lambda}$ and
$x_{j}\rightarrow x^{\prime}$ give
\beq
N(x)=N_{0}\int^{x}_{x_{\rm{min}}}
\frac{dx^{\prime}}{\lambda}{\rm{e}}^{\frac{(x-x^{\prime})}
{\xi_{e}}}
= N_{0}\frac{\xi_{e}}{\lambda}\left[
\rm{e}^{\frac{(x-x_{\rm{min}})}{\xi_{e}}}-
1\right].
\label{eq:nofx}
\eeq
The physical interpretation of the limits of integration
is as follows:
$x_{\rm{min}}$ is the smallest depth at which an
initiated subshower is
still cascading at the depth $x.$  Subshowers initiated
at depths $<x_{\rm{min}}$ will have already ranged out
by the point $x$ \cite{foot6}, and
so they cannot contribute to the particle
number at $x$ and are excluded from the integral.

From the geometry of fig.~(\ref{fig:monoshower}),
using eqs.~(\ref{eq:xmax}) and (\ref{eq:Ej})
and taking $j\rightarrow
\frac{x}{\lambda}$, we can deduce
\beq
x_{\rm{min}}(x)=x-X_{\rm{max}}^{j}= x-
\xi_{e} \ln\left(\frac{\eta(1-\eta)^{\frac{x_{\rm{min}}}
{\lambda}}E_{0}}{(1-\eta)N_{0}E_{c}} \right)\,.
\eeq
Solving for $x_{\rm{min}}$ gives
\beq
x_{\rm{min}}=\frac{x-\xi_{e}\ln\left(\frac{\eta E_{0}}
{(1-\eta) N_{0}E_{c}}\right)}
{1+\frac{\xi_{e}}{\lambda}\ln(1-\eta)}\,.
\eeq
Using $\eta\ll 1,$ so $\ln(1-\eta)\simeq -\eta$
and $\frac{\eta}{1-\eta}\simeq \eta,$ this result simplifies to
\beq
x_{\rm{min}}=\frac{x-X}{1-\frac{\eta\xi_{e}}{\lambda}}\,.
\label{eq:xmin}
\eeq
Here and in the remainder of this section
we define $X\equiv X_{\rm{max}}^{0}=\xi_{e}\ln
\left(\frac{\eta E_{0}}{N_{0}E_{c}}\right),$ which is the
depth of the first subshower maximum 
(coming from a  monopole with its full kinetic energy $E_{0}$).
Expression (\ref{eq:xmin}) is formally singular at 
$\eta =\lambda/\xi_e\ll 1$, but 
for monopoles this singular point is not relevant.


For a monopole passing through matter, 
$\frac{\eta\xi_e}{\lambda}\ll 1$,
so we can write eq.~(\ref{eq:xmin}) as
\beq
x-x_{\rm{min}}=X-\frac{\eta\xi_e}{\lambda} x\,.
\eeq
Substituting the above into eq.~(\ref{eq:nofx}) gives
the quasi steady--state shower size
\beq
N(x)
\simeq
\frac{\xi_{e}}{\lambda}
\exp\left(\frac{X}{\xi_{e}}\right)
\label{eq:nofx2}
\eeq

Notice that the $x$--dependence has vanished, as is appropriate
for a steady--state phenomenon.
Using the definition of $X$ above, and 
the useful relation between the continuum and discrete
expressions for energy loss
\beq
\frac{\eta E}{\lambda}=\left|\frac{dE}{dx}\right|\,,
\label{eq:stopapprox}
\eeq
we rewrite (\ref{eq:nofx2}) as
\beq 
N(\gamma)=\frac{\xi_{e}}{E_{c}}
\left|\frac{dE_{\rm{pair}}}{dx}
\right|_{\eta_{\rm{crit}}<\eta}\,,
\label{eq:nofx3}
\eeq
where $\eta_{\rm{crit}}\equiv \frac{200\rm{MeV}}{E}$.

\begin{figure}[c]
\centerline{\hbox{
\epsfxsize=210pt \epsfbox{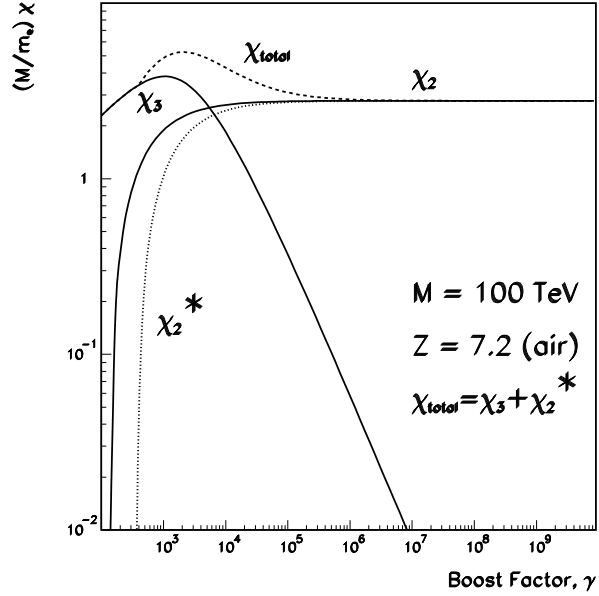}
}}
\caption[Monopole-Induced Pair Production For $\eta>\eta_{\rm{crit}}$.]
{Electron-positron pair production of an $M=100$ TeV relativistic
monopole in air where the inelasticity is restricted
to $\eta>\eta_{\rm{crit}}\equiv\frac{200\rm{MeV}}{\gamma M}$.}
\label{fig:pairint2}
\end{figure}
We next turn to some details of the shower energy.
In fig.~(\ref{fig:pairint2}) we plot the dominant terms
for energy loss to electron--pair production, $\chi_{2}$
and $\chi_{3}.$  For $\gamma \gsim 10^{4}$, pair production
is dominated by $\chi_{2},$ which describes slow pairs
in the no screening limit.  For shower development we
must restrict
the inelasticity to $\eta>\eta_{\rm{crit}}\equiv\frac{200{\rm MeV}}{E}$.
The plot $\chi_{2}^{*}$ shows the effect of this restriction.
Notice that $\chi_{2}^{*}\simeq\chi_{2}$ for $\gamma \gsim 10^{5},$
so we are justified in using eq.~(\ref{eq:pairstop2}) without the
$\eta$ restriction to compute the shower
size when $\gamma \gsim 10^{5}.$
In the evaluation of eq.~(\ref{eq:nofx3}) we
use a parameterization of $\xi_{e}$ in different
media (good to $\lsim 2.5\%$) given
by \cite{Partdata}
\beq
\xi_{e}=\frac{716.4\; A}
{Z(Z+1)\;\ln\left(\frac{287}{\sqrt{Z}}\right)}
\;\rm{cm}^{-2}
\eeq
and a fit to the data \cite{Rossi} for $E_{c}:$
\beq
\frac{E_{c}}{\rm{MeV}}=\left\{\ba{ll}\frac{610}
{Z+1.24} & \;\;\rm{for}\;\rm{solids}\;\rm{and}
\;\rm{liquids}\\\frac{710}
{Z+0.92} & \;\;\rm{for}\;\rm{gases}\ea\right\}.
\eeq

The contribution of the photonuclear process to the 
{\it{electromagnetic}} shower is indirect.  The photonuclear 
interaction injects hadrons into the monopole shower.
A subshower initiated by a high energy 
hadron will produce $\pi^{0}$'s as secondaries, which each decay 
to 2 $\gamma$'s.  If these $\gamma$'s have $E > E_{c},$ 
they may initiate an electromagnetic shower.  
So, only the largest inelasticity fraction of the energy lost 
via the photonuclear interaction contributes to the
electromagnetic shower in the end.
Given this fact, it is reasonable to 
assume that pair production 
alone provides a lower bound to the electromagnetic
shower size and that the pair production plus photonuclear
interaction provides an upper bound.  


\begin{figure}[b]
\centerline{\hbox{
\vspace{6.5 in}
\epsfxsize=210pt \epsfbox{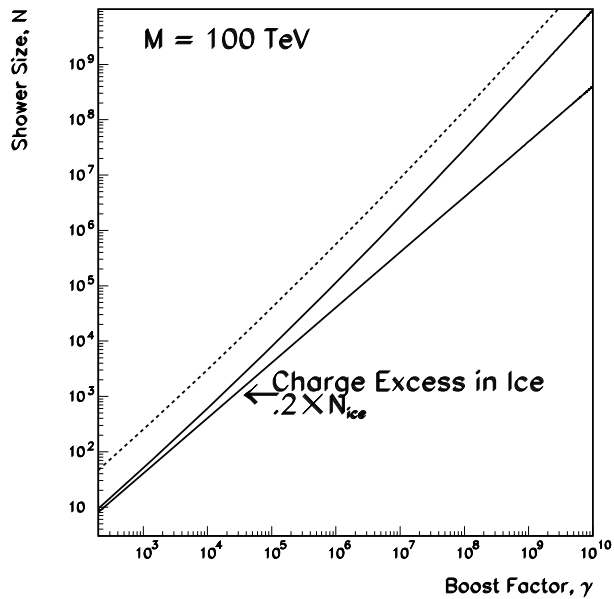}
}}
\caption{The monopole-induced shower 
in ice for a monopole of mass $100$ TeV.  The shower
size is the total number of electron, positrons, and
photons.  The dashed line is the total shower from pair
production and the photonuclear interaction.
The solid lines show the electric charge excess 
(roughly 20\% of the shower size) for pair production alone
($\propto \gamma$)
and pair production plus photonuclear ($\propto \gamma^{1.28}$).}
\label{fig:shower}
\end{figure}


\subsubsection{Lateral Shower Profile}

For monopole-induced showers the lateral profile
is greatly simplified in comparison to hadronic
primaries.  As shown previously the monopole-induced
shower is approximately constant,
being continuously regenerated by the monopole 
as the lower energy particles range out.
For our purposes in the following
section it is sufficient to assume that the lateral
profile is uniform out to a lateral cutoff given by the
Moli\`{e}re radius
\beq
R_{\rm{M}}=7.4\,\frac{\rm{g}}{\rm{cm}^{2}}\;
(\frac{\xi_{e}}{35 {\rm g/cm^2}})
\,(\frac{100 \rm{MeV}}{E_{c}}).
\label{eq:moliere}
\eeq
As defined, the Moli\`{e}re radius is
independent of the incident monopole energy, being
determined only by the
spread of low energy particles resulting from
multiple Coulomb scattering.  Within a
distance $R_{\rm{M}}$ of the monopole path, 
$\sim$~90~\% of the shower particles will 
be found \cite{Partdata}.


\section*{Acknowledgements}

We thank the Aspen Center for Physics for the environment that allowed
this collaboration to begin.  The Vanderbilt group thanks A. Berera,
A. Kusenko, G. Medina--Tanco, J. Ralston for discussion, and 
especially David Seckel for discussions and mind altering criticism.
P. Biermann would like to thank Drs. H. Kang, P. Kronberg,
C. Quigg and D. Ryu for extensive discussions on monopoles and
magnetic fields in the cosmos.
This work was supported in part by the U.S.
Department of Energy grant no. DE-FG05-85ER40226,
the Vanderbilt University Research Council,
and the NASA/Tennessee Space Grant Consortium.


\begin{thebibliography}{99}

\bibitem{Kibble}
T.~W.~Kibble,
Phys.\ Rept.\  {\bf 67}, 183 (1980), and references therein.
%
\bibitem{LPi} Late--term phase transitions may temporarily
break $U_{EM}(1),$ facilitating monopole annihilation.  However,
these models are rather contrived.  See
P.~Langacker and S.--Y.~Pi, Phys. Rev. Lett. {\bf{45,}} 1 (1980);
T.H.~Farris, T.W.~Kephart, T.J.~Weiler, and T.C.~Yuan,  
Phys. Rev. Lett. {\bf{68,}} 564 (1992);
V.V.~Dixit and M.~Sher,
Phys. Rev. Lett. {\bf{68,}} 560 (1992).
%
\bibitem{refs} G.~Giacomelli, et al., Magnetic Monopole
Bibliography, DFUB 9/98 (1998).
%
\bibitem{Parker} E. N. Parker, Astrophys. J. {\bf 160}, 383 (1970);
{\it {ibid.}\ }, {\bf 163}, 225 (1971);
{\it {ibid.}\ },{\bf 166}, 295 (1971).
%
Based on more speculative hypotheses, more stringent bounds have 
been proposed for the galactic flux by
M.S. Turner, E.N. Parker, and T. J. Bogdan, Phys. Rev. {\bf D26}, 1296 (1982);
and for the extragalactic flux by
Y. Rephaeli and M. S. Turner, Phys. Lett. {\bf 121B}, 115 (1983);
F. Adams et al., Phys. Rev. Lett. {\bf 70}, 2511 (1993);
M.J. Lewis, K. Freese, and G. Tarle, [astro-ph/9911095].  It is
not clear to us that any of these bounds apply to a relativistic
flux of monopoles; a relativistic monopole traveling through a
plasma may actually create magnetic fields in its wake.  
%
\bibitem{PorterGoto} N.A. Porter, Nuovo Cim. {\bf 16}, 958 (1960);
E. Goto, Prog. Theo. Phys. {\bf 30}, 700 (1963).
%
\bibitem{KW96}  T.W. Kephart and T.J. Weiler,
Astropart. Phys. {\bf{4}} 271 (1996);
Nucl. Phys. (Proc. Suppl.)  {\bf 51B}, 218 (1996).
%
\bibitem{tHooft}  G. 't Hooft, Nucl. Phys. {\bf{B79,}} 276 (1974);
A. Polyakov, JETP Lett. {\bf{20,}} 194 (1974).
%
\bibitem{HK}
For a review and references see: M.~B.~Hindmarsh and T.~W.~Kibble,
Rept.\ Prog.\ Phys.\  {\bf 58}, 477 (1995)
[hep-ph/9411342].
%
\bibitem{DeRujula} A. De Rujula, Nucl. Phys. {\bf B435}, 257 (1995).
\bibitem{foot1}
Due to the smallness of the monopole's Thomson cross--section
for monopole masses of interest, \\
$\sigma_T\sim 2\times 10^{-43}(10^{10}{\rm GeV}/M)^2 {\rm cm}^2$,
the mean free path for a monopole to scatter on the
cosmic background radiation greatly exceeds the Hubble size
of the Universe \cite{KW96}.
\bibitem{SK}
S.~F.~King and Q.~Shafi,
Phys.\ Lett.\  {\bf B422}, 135 (1998)
[hep-ph/9711288].
%
\bibitem{Hong} D.K.~Hong, J.~Kim, J.E.~Kim, and K.S.~Soh,
Phys.\ Rev.\  {\bf D27}, 1651 (1983).
%
\bibitem{Dutta}
N.~G.~Deshpande, B.~Dutta and E.~Keith,
Nucl.\ Phys.\ Proc.\ Suppl.\  {\bf 52A}, 172 (1997)
[hep-ph/9607307];
Phys.\ Lett.\  {\bf B388}, 605 (1996)
[hep-ph/9605386];
Phys.\ Lett.\  {\bf B384}, 116 (1996)
[hep-ph/9604236].
%
\bibitem{FLFK}
P.~H.~Frampton and B.~Lee,
Phys.\ Rev.\ Lett.\  {\bf 64}, 619 (1990);
P.~H.~Frampton and T.~W.~Kephart,
Phys.\ Rev.\  {\bf D42}, 3892 (1990).
%
\bibitem{Mald}  M. Maldacena, Adv. Theor. Math. Phys.
{\bf D42}, 231 (1998).
%
\bibitem{PHF}
P.~H.~Frampton,
Phys.\ Rev.\  {\bf D60}, 121901 (1999)
[hep-th/9907051].
%
\bibitem{PHF2}
P.~H.~Frampton,
Phys.\ Rev.\  {\bf D60}, 085004 (1999)
[hep-th/9905042];
P.~H.~Frampton and T.~W.~Kephart,
[hep-th/9912028].
%
\bibitem{Dienes}
K. R. Dienes, E. Dudas, and T. Gherghetta,
Phys. Lett. {\bf B436}, 55 (1998)
[hep-ph/9803466].
%
\bibitem{Dimop}
I.Antoniadis, N. Arkani-Hamed, S. Dimopoulos, and
G. Dvali, Phys. Lett. {\bf B436}, 257 (1998).
[hep-ph/9804398]
%
\bibitem{Harvey2} C.L. Gardner and J.A. Harvey,
Phys. Rev. Lett. {\bf{52,}} 879 (1984).
%
\bibitem{Tanmay2}
T.~Vachaspati,
Phys.\ Rev.\ Lett.\  {\bf 76}, 188 (1996)
[hep-ph/9509271];
H.~Liu and T.~Vachaspati,
Phys.\ Rev.\ {\bf D56}, 1300 (1997)
[hep-th/9604138].
%
\bibitem{Goldhaber}
A.~S.~Goldhaber,
Phys.\ Rept.\  {\bf 315}, 83 (1999)
[hep-th/9905208].
%
\bibitem{Kron}
For a review see P.~P.~Kronberg,
Rept.\ Prog.\ Phys.\  {\bf 57}, 325 (1994); galactic cluster
fields are updated in T.E.~Clark, P.P.~Kronberg, and H.~Boehringer,
to appear in Ap. J. Lett., [astro-ph/0011281].
%
\bibitem{vortons}
E.~Huguet and P.~Peter,
Astropart. Phys. {\bf{12}}, 277 (2000)
[hep-ph/9901370].
%
\bibitem{Ryu97}  D. Ryu, H. Kang, and
P.L. Biermann, Astron. \& Astrophys. {\bf335}, 19 (1998)
[astro-ph/9803275].
%
\bibitem{Kim89}  K.-T. Kim et al., Nature {\bf 341}, 720 (1989).
%
\bibitem{Beck96}  R. Beck, Ann. Rev. Astron. \& Astrophys.  {\bf 34}, 155
(1996).
%
\bibitem{Kronberg}
P.P. Kronberg, P.L. Biermann and F.R. Schwab,
Astrophys. J. {\bf 291}, 693 (1985).
%
\bibitem{Kellerman}
K.I. Kellermann and I.I.K Pauliny-Toth, Ann. Rev. A\&A
{\bf 19}, 373 (1981).
%
\bibitem{Ensslin97}  T.A. En{\ss}lin, et al., Astrophys. J.  {\bf 477},
560 - 567 (1997).
%
\bibitem{Plaga}  R. Plaga, {\it Nature} {\bf 374}, 430 (1995).
\bibitem{E6} A third possible type of GUT monopoles is truly
``color blind.''  For example, an
$E_{6}$ GUT with breaking scheme \\
$E_{6}\rightarrow SU_{C}(3)\times SU_{L}(3)\times
SU_{R}(3) \rightarrow SU_{C}(3)\times SU_{L}(2)\times
U_{Y}(1)$ has monopoles with no hadronic interaction whatsoever.
\bibitem{large}
In cgs units, $\alpha=e^2/4\pi$, $\alpha_M=g^2/4\pi$,and $g\,e=2\pi$.
In Heaviside-Lorentz units,
$\alpha=e^2$, $\alpha_M=g^2$,and $g\,e=\half$.
In either set of units, $g=e/2\alpha$ and $\alpha_M=1/4\alpha$.
\bibitem{foot2}
Recent progress in quantizing the interacting
monopole can be found in
M. Blagojevic and P. Senjanovic, Phys. Rept. {\bf 157}, 233 (1988);
L. Gamberg and K. A. Milton, hep-ph/9910526.
\bibitem{Giac1}  G. Giacomelli, in:
{\it{Theory and Detection of Magnetic Monopoles
in Gauge Theories,}} ed. N. Craigie p.407
(Singapore: World Scientific Publishing Co., 1986).
\bibitem{Ahlen1}  S.P. Ahlen, Rev. Mod. Phys. {\bf{52,}}
121 (1980).
\bibitem{LL} L.D.~Landau and E.M.~Lifshitz,
{\it{Electrodynamics of Continuous Media,}}
(Oxford: Pergamon Press, 1984).
%
\bibitem{Effphoton} L.D.~Landau and E.M.~Lifshitz,
{\it{Relativistic Quantum Mechanics,}}
(Oxford: Pergamon Press, 1984).
%
\bibitem{Sternheimer} R.M. Sternheimer, R.F. Peierls,
Phys. Rev. B {\bf{3,}} 3681 (1971).
%
\bibitem{Kelner} S.R. Kel'ner and Yu.D. Kotov,
Sov. Jour. Nucl. Phys. {\bf{7,}} 237 (1968); S.R.
Kel'ner, Sov. Jour. Nucl. Phys. {\bf{5,}} 778 (1967).
%
%
%
%
\bibitem{Jackson}  J. D. Jackson,
{\it{Classical Electrodynamics}}, (New York: John
Wiley and Sons, 1975).
%
\bibitem{foot3}
For bremsstrahlung, screening of the 
nuclear charge by the atomic electrons
is large when \cite{Jackson}
\beq
\gamma > \frac{192M}{m_{e}Z^{\frac{1}{3}}}
\simeq \frac{3\times 10^{10}}{Z^{\frac{1}{3}}}
\left(\frac{M}{100\rm{TeV}}\right)\,.
\label{eq:screening}
\eeq
Even for the lightest monopoles
which we consider ($40$ TeV),
eq. (\ref{eq:screening}) shows that screening
corrections are expected to be small.
%
%
\bibitem{Reno} S.I.~Dutta, M.H.~Reno, I.~Sarcevic, and D.~Seckel,
Phys. Rev. D {\bf{63,}} 094020 (2001) [hep-ph/0012350].
%
\bibitem{Heitler}  W. Heitler, {\it{The Quantum
Theory of Radiation,}} (Oxford: Clarendon Press, 1954).
%
\bibitem{Gaisser} T.K.~Gaisser,
{\it{Cosmic Rays and Particle Physics,}}
(Cambridge: Cambridge University Press, 1990);
or A. M. Hillas, {\sl Ann. Rev. Astron. Astrophys.} {\bf 22}, 425 (1984).
%
\bibitem{Sokolsky} P. Sokolsky, {\it{Introduction to
Ultrahigh Energy Cosmic Ray Physics}}, (New York:
Addison-Wesley Publishing Company, 1989).
\bibitem{foot4} The particle
doubling length $R$ can be lengthened significantly
by the Landau-Pomeranchuk-Migdal effect
\cite{Klein,LPM} for photon or electron
energies $E \gsim E_{\rm{LPM}}.$
The value of $E_{\rm{LPM}}$ is dependent
on the material and some typical values are
$E_{\rm{LPM}}=2.2$ TeV for lead, $E_{\rm{LPM}}=139$ TeV
for water, and $E_{\rm{LPM}}=117$ PeV for sea level air
\cite{Klein}.   This model assumes that
$E \ll E_{\rm{LPM}}$ which is generally the case
for $e^{\pm}$ and $\gamma$ originating from
monopole--nucleus interactions.
\bibitem{Klein}  S. Klein, [astro-ph/9712198]; ibid, 
Rev. Mod. Phys. {\bf{71,}} 1501 (1999).
%
\bibitem{LPM}  L.D. Landau and I.J. Pomeranchuk, in
{\it{The Collected Papers of L.D. Landau,}}
(Oxford: Pergamon Press, 1965);
A.B. Migdal, Phys. Rev. {\bf{103,}} 1811 (1956).
%
\bibitem{Zas}  E. Zas, F. Halzen, and T. Stanev,
Phys. Rev. D {\bf{45,}} 362 (1992).
%
\bibitem{Clay} R. Clay and P. Gerhardy, Australian J. Phys.
{\bf{35,}} 59 (1982).
%
\bibitem{Fermi}  E. Fermi, Phys. Rev.
{\bf{ 57,}} 485 (1940).
%
\bibitem{Mpolzn} R. Hagstrom, {\sl Phys. Rev. Lett.} {\bf 35}, 1677 (1975).
%
%
\bibitem{Baikal} V.A. Balkanov et al.\ (Baikal Collaboration),
$26^{\rm th}$ International Cosmic Ray Conf.\ (ICRC99),
Salt Lake City, UT, Aug.\ 17-25, 1999, paper H.E.5.3.04.
%
\bibitem{AMANDA} P. Niessen et al.\ (AMANDA Collaboration),
$26^{\rm th}$ International Cosmic Ray Conf.\ (ICRC99),
Salt Lake City, UT, Aug.\ 17-25, 1999, paper H.E.5.3.05.
%
\bibitem{MACRO} B.C. Choudhary et al.\ (MACRO Collaboration),
$26^{\rm th}$ International Cosmic Ray Conf.\ (ICRC99),
Salt Lake City, UT, Aug.\ 17-25, 1999, paper H.E.5.3.02.
%
\bibitem{Halzen} F.~Halzen, private communication.
%
\bibitem{foot7}
The charge excess
results from a number of processes involving the
shower interacting with
the medium.  These include knock-on electrons,
Compton scattering, and positron annihilation.
\bibitem{Askaryan} G.A.~Askar'yan, Sov. Phys. JETP
{\bf{14}}, 441 (1962); {\it{ibid.}}, {\bf{48}}, 988 (1965);
M.A.~Markov and I.M.~Zeleznykh, Nucl. Instrum. Methods A
{\bf{248}}, 242 (1986).
%
\bibitem{Frichter} G.M. Frichter, J.P. Ralston, and
D.W. McKay, Phys. Rev. D {\bf{53,}} 1684 (1996).
\bibitem{MACRO2}
Calculations of the energy loss of monopoles and dyons passing
through the Earth and in scintillators, streamer tubes,
nuclear track detectors is given in 
J.~Derkaoui, Astro. Part. Phys., {\bf{9,}} 173 (1998);
Astro. Part. Phys., {\bf{10,}} 339 (1999). 
\bibitem{foot11}
For example, the energy threshold for Fly's Eye is
$\sim 10^{17.3}$ eV \cite{Fly} which renders the Eye blind
to $l$-monopoles, except for those with
the smallest allowed mass at the greatest allowed kinetic energies.
%
\bibitem{Fly} D. J. Bird et al. (Fly's Eye Collab.),
Astrophys. J. {\bf 424}, 491 (1994); {\sl ibid.} {\bf 441}, 144
(1995).
\bibitem{Lay} T. Lay and T.C. Wallace,  {\it{Modern Global
Seismology}}, (New York: Academic Press, 1995).
\bibitem{Geo} Properties of the Solid Earth, in Rev.
Geophys. {\bf{33,}} (1995); also at the URL:
http://earth.agu.org/revgeophys.
\bibitem{foot8}
This idea has been exploited in neutrino
physics as neutrinos are sufficiently weakly interacting
to pass through the earth largely unimpeded
for neutrino energies $\lsim 10^{15}\;\rm{eV}.$
\bibitem{Anderson} A.M. Dziewonski and D.L. Anderson,
Phys. Earth Planet. Inter. {\bf{25,}} 297 (1981).
\bibitem{foot9}
In principle, the eight shell model could
be used in our calculations but it provides
more detail than is needed here.  The two shell
approximation includes the gross features necessary
to show tomography with monopoles.
\bibitem{Kraus} K.B. Krauskopf and D.K. Bird,
{\it{Introduction to Geochemistry}}
(New York: McGraw-Hill Publishers, 1995).
\bibitem{foot10}
The angular resolution of earth tomography with
monopoles is, in principle, limited by the multiple scattering
of the monopole through small angles.  However, the large
monopole mass renders this angular uncertainty negligibly small.
%
\bibitem{ringberg} T.J. Weiler, hep-ph/9910316, 
based on private communication from A.A. Watson.
%
\bibitem{Minertia} R.N. Mohapatra and S. Nussinov,
Phys. Rev. {\bf D57}, 1940 (1998);
in I.F.M. Albuquerque, G. Farrar, and E.W. Kolb,
Phys. Rev. {\bf D59}, 015021 (1999) it is noted that a baryon mass
above 10 GeV produces a noticeably different shower profile,
and a baryon mass above 50 GeV is so different as to be ruled out.
\bibitem{foot12}
In the other extreme, it has been proposed \cite{BV} that the origin
of the superGZK events are the gauge bosons radiated by cosmic
monopole--string networks when the stretched strings relax.
\bibitem{BV} V. Berezinsky, X. Martin, and A. Vilenkin,
Phys. Rev. {\bf D56}, 2024 (1997);
V. Berezinsky and A. Vilenkin, Phys. Rev. Lett. {\bf 79}, 5202 (1997);
V. Berezinsky, P. Blasi, and A. Vilenkin
[astro-ph/9803271].
%
\bibitem{GR} More exactly, $\sum_{n=1}^N =\ln N +\gamma_E 
+ \frac{1}{2N} + {\cal O}(\frac{1}{N^2})$,
where $\gamma_E = 0.577\dots$ is Euler's constant;
this formula can be found on p.\ 2 of I.S. Gradshteyn and I.M. Ryzhik,
{\it Tables of Integrals, Series, and Products}, 
(New York: Academic Press\, 1980).
\bibitem{foot6}
We neglect particles after their energies fall below $E_c$.
Omission of this exponentially--decaying tail results in an
underestimate of
the true particle number by a factor of order a few.
This makes a small difference as far as the coherent--shower
Cherenkov emission is concerned.
It makes more difference to the longitudinal profile, and therefore to
the fluorescence signal.  However, we show in section (\ref{sec:fluor})
that the fluorescence signal from a monopole is weak, probably too weak
to be measured.
\bibitem{foot13}
Our cutoff on shower development for particles 
with energy below $E_c = 200$~MeV is conservative.  For example,
an $e^+e^-$ pair produced by a 50~MeV photon yields 
25~cm of tracklength in ice, along which the pair will
generate Cherenkov light, delta rays, and further ionization.
\bibitem{Partdata} Particle Data Group,
Phys. Rev. D {\bf 50}, 1173 (1994).
%
\bibitem{Rossi} B. Rossi, {\it{High-Energy Particles}},
(New York: Prentice-Hall, 1952).
%
%
\end{thebibliography}
\end{document}